\documentclass[twocolumn]{aastex631}

\shorttitle{The Dynamical Consequences of a Super-Earth in the Solar
  System}
\shortauthors{Stephen R. Kane}

\begin{document}

\title{The Dynamical Consequences of a Super-Earth in the Solar
  System}

\author[0000-0002-7084-0529]{Stephen R. Kane}
\affiliation{Department of Earth and Planetary Sciences, University of
  California, Riverside, CA 92521, USA}
\email{skane@ucr.edu}


\begin{abstract}

Placing the architecture of the Solar System within the broader
context of planetary architectures is one of the primary topics of
interest within planetary science. Exoplanet discoveries have revealed
a large range of system architectures, many of which differ
substantially from the Solar System model. One particular feature of
exoplanet demographics is the relative prevalence of super-Earth
planets, for which the Solar System lacks a suitable analog,
presenting a challenge to modeling their interiors and atmospheres.
Here we present the results of a large suite of dynamical simulations
that insert a hypothetical planet in the mass range 1--10~$M_\oplus$
within the semi-major axis range 2--4~AU, between the orbits of Mars
and Jupiter. We show that, although the system dynamics remain largely
unaffected when the additional planet is placed near 3~AU, Mercury
experiences substantial instability when the additional planet lies in
the range 3.1--4.0~AU, and perturbations to the Martian orbit
primarily result when the additional planet lies in the range
2.0--2.7~AU. We further show that, although Jupiter and Saturn
experience relatively small orbital perturbations, the angular
momentum transferred to the ice giants can result in their ejection
from the system at key resonance locations of the additional
planet. We discuss the implications of these results for the
architecture of the inner and outer solar system planets, and for
exoplanetary systems.

\end{abstract}

\keywords{astrobiology -- planetary systems -- planets and satellites:
  dynamical evolution and stability}


\section{Introduction}
\label{intro}

The discovery of numerous exoplanets over the past several decades has
allowed statistical studies of planetary architectures that place the
Solar System within a broader context
\citep{limbach2015,ford2014,martin2015b,winn2015,horner2020b,kane2021d}. The
detected multi-planet systems exhibit a remarkable diversity of
architectures that differ substantially from the Solar System
\citep{hatzes2016d,he2019}. Examples include compact systems, such as
the Kepler-11 \citep{lissauer2011a,gelino2014} and TRAPPIST-1
\citep{gillon2016,gillon2017a,luger2017b,agol2021,kane2021c} systems,
demonstrating that planetary systems may harbor multiple planetary
orbits with long-term stability close to the host star. Additionally,
exoplanetary systems reveal a wide range of orbital eccentricities
\citep{shen2008c,kane2012d,vaneylen2015}, as extreme as 0.96 for the
planet HD~20782b \citep{kane2016b,udry2019}, and evidence for past
significant planet-planet interactions
\citep{juric2008b,kane2014b,carrera2019b,bowler2020a}. Studying the
statistics of exoplanet demographics offers the opportunity to provide
important insights into the question of how typical our Solar System
architecture and evolution is compared with other planetary systems
\citep{limbach2015,martin2015b}.

One of the current main challenges to placing the Solar System within
the broad planetary formation context are planetary sizes/masses that
are not present within our system. In particular, exoplanet
discoveries have included a large population of super-Earths
\citep{valencia2007b,leger2009,howard2010b,bonfils2013a} and
mini-Neptunes \citep{barnes2009a,lopez2014,nielsen2020b}, planets
whose size places them within the Solar System's notable size gap
between Earth and Neptune. The local dearth of this category of
planets results in a lack of in-situ data, truncating models of their
possible composition and formation differences compared with the Solar
System ice giants \citep{owen2017c,lee2019}. The factors that
contributed to the $\sim$1--4~$R_\oplus$ radius gap within the Solar
System are numerous, but predominantly is due to early migration of
the giant planets
\citep{gomes2005b,morbidelli2005,walsh2011c,raymond2014a,nesvorny2018c},
a process that may also have limited the growth of Mars
\citep{bromley2017,clement2018,clement2021f}. Even so, it is useful to
investigate the dynamical consequences of additional planetary mass
within the Solar System, in order to constrain current formation
theories and study the implications for general planetary system
architectures \citep{lissauer2001c,nesvorny2012c}.

In this paper, we provide the results of a dynamical study that places
an additional terrestrial planet in the mass range 1--10~$M_\oplus$
and semi-major axis range 2--4~AU within the current Solar System
architecture. Section~\ref{why} summarizes the evidence for giant
planet migration and the reasons why the Solar System does not have a
super-Earth between the orbits of Mars and Jupiter. The details of the
executed suite of dynamical simulations are described in
Section~\ref{alternate}, together with specific examples and global
results for each of the Solar System planets. Section~\ref{con}
discusses the implications of these results for the inner planets,
outer planets, and exoplanetary system architectures. We provide
concluding remarks and suggestions for further work in
Section~\ref{conclusions}.


\section{Why is There No Super-Earth in the Solar System?}
\label{why}

The architecture of the Solar System planetary orbits exhibits a
substantial gap between the orbits of Mars and Jupiter, populated by a
vast reservoir of asteroids. These features of the Solar System
architecture provide compelling evidence of past interactions, and
numerous chellanges remain for Solar System formation models
\citep{morbidelli2016b}. The gravitational interactions of Jupiter
with the material within the asteroid belt is well known, particularly
those related to mean motion resonances (MMR) and the Kirkwood gaps
\citep{dermott1981b,wisdom1983a,morbidelli1989b,gladman1997b}. Earlier
work strongly suggested that the total mass of the asteroid belt has
been significantly depleted since the formation of the planets
\citep{bottke2005b}, whereas more recent work indicates that the
present asteroid belt is consistent with relatively empty primordial
origins \citep{raymond2017a}.

However, it is generally agreed that the Solar System giant planets
have undergone migration processes that have altered their locations
through time relative to where they formed. Models that describe the
giant planet migration range from the early ``Grand Tack'' model of
orbital evolution for Jupiter and Saturn, in which Jupiter migrated
inward to approach the current orbit of Mars before moving back
outward \citep[e.g.,][]{walsh2011c,raymond2014a,nesvorny2018c}, to the
late and chaotic interactions of the ``Nice Model''
\citep[e.g.,]{gomes2005b,morbidelli2005}. The formation of the four
terrestrial planets and associated accretion models has been studied
in detail \citep{raymond2009c,lykawka2019}, for which the migration of
the giant planets played a major role in sculpting their eventual size
and orbits \citep{kaib2016a,nesvorny2021a}. Indeed, the size of Mars
and the distribution of asteroids within the main belt are essential
clues to the terrestrial planet formation processes that occurred
\citep{demeo2014b,izidoro2015c}.

The asteroid belt further provides plentiful evidence for the
migration of the Solar System giant planets and their gravitaional
influence on the architecture of the inner Solar System
\citep{morbidelli2010a,deienno2016a,izidoro2016,clement2020a,kane2020e}. In
particular, planetary migration possibly played a major role in
depletion of material between the orbits of Mars and Jupiter
\citep{clement2019a,pirani2019a}, including the effect of secular
resonances with the giant planets as they migrated inward and outward
\citep{minton2011}. Furthermore, exoplanet discoveries have revealed
an abundance of super-Earths and mini-Neptunes that are relatively
close to their host stars \citep{winn2015}. The lack of any such
planets interior to the orbit of Mercury has thus become a matter of
some intrigue for Solar System formation models that address the
architecture of the inner Solar System
\citep{morbidelli2012a,clement2022a}. Explanations for the relative
dearth of inner Solar System planets include outward migration of
planet forming material due to the solar wind \citep{spalding2018c}, a
narrow annulus of building material \citep{hansen2009c},
dynamical/collisional destruction of inner planets \citep{volk2015},
and the presence and orbital migration of the giant planets
\citep{batygin2015b,hansen2017b}. Indeed, as dynamical models of the
early Solar System continue to evolve, the evidence increasingly
suggests that it was the powerful combination of multiple giant
planets and their migration processes that prevented planet formation
in the region of the asteroid belt, and ensured that the inner Solar
System did not harbor a super-Earth planet.


\section{An Alternate Architecture}
\label{alternate}

Here we describe the details of the dynamical simulations conducted
for our study, and the overall results for each of the Solar System
planets.


\subsection{Dynamical Simulation}
\label{sim}

\begin{figure}
  \includegraphics[angle=270,width=8.5cm]{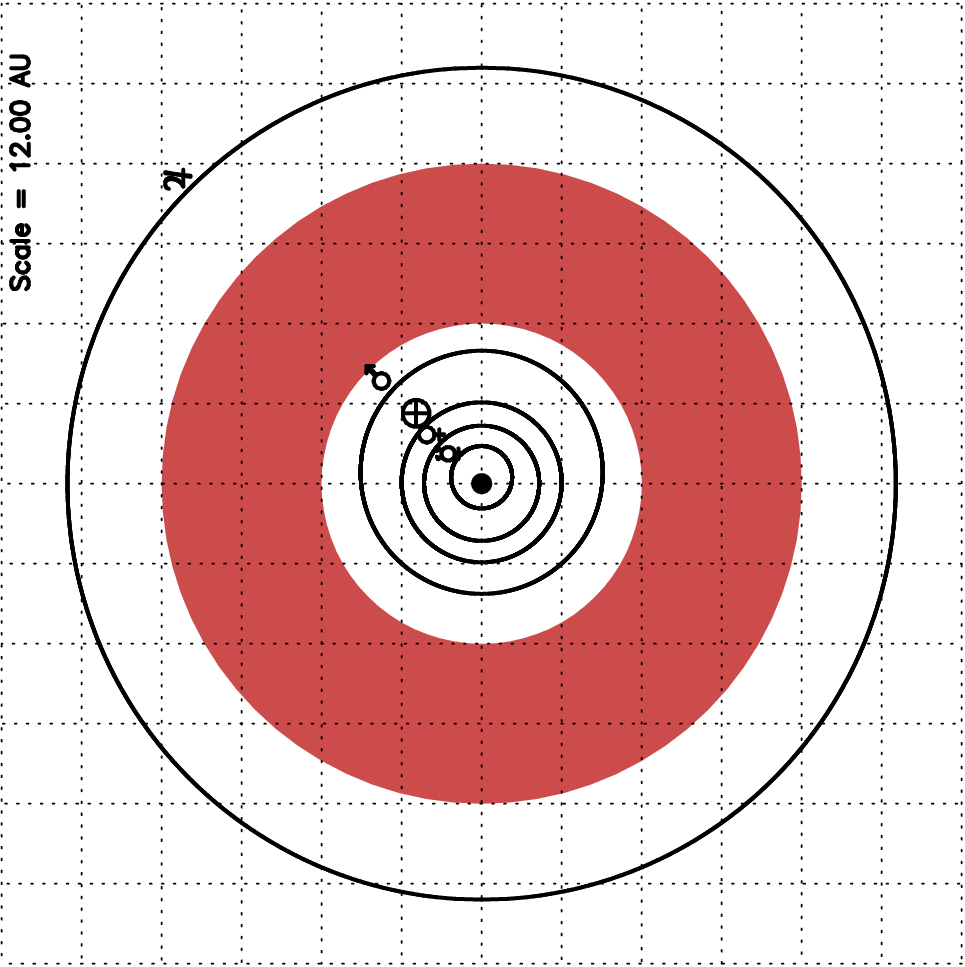}
  \caption{Top-down view of the Solar System planetary orbits (shown
    as solid lines) out to the orbit of Jupiter. The red shaded region
    indicates the locations in which a planet in the mass range
    1--10~$M_\oplus$ was added to the Solar System architecture. The
    dotted lines show a grid with 1~AU resolution, and the full figure
    scale is 12~AU along one side.}
  \label{fig:orbits}
\end{figure}

The N-body integrations for the analysis reported here were performed
with the Mercury Integrator Package \citep{chambers1999}, where the
methodology was similar to that described by
\citet{kane2014b,kane2016d,kane2019c}. The orbital elements for the
major planets were extracted from the Horizons DE431 ephemerides
\citep{folkner2014}; minor planets were not included in the
simulation. The simulations used a hybrid symplectic/Bulirsch-Stoer
integrator with a Jacobi coordinate system to provide increased
accuracy for multi-planet systems
\citep{wisdom1991,wisdom2006b}. Given the relatively large semi-major
axes involved in our simulations, General Relativty was not calculated
in the evolution of the planetary orbits. Dynamical simulations
conducted by \citet{lissauer2001c} provided a dynamical simulation
that added mass to Ceres and several other asteroids, finding
dynamically stable configurations for a subset of the
simulations. Here, we provide a more general case by systematically
exploring the parameter space of planetary mass and semi-major axis of
the added body. Specifically, the simulations described here add a
planet with mass range 1--10~$M_\oplus$, in steps of 1~$M_\oplus$. The
additional planet was placed at various starting locations in a
circular orbit, coplanar with Earth's orbit, and with a semi-major
axis range 2--4~AU, in steps of 0.01~AU, as shown in
Figure~\ref{fig:orbits}. This resulted in several thousand
simulations, where each simulation was allowed to run for $10^7$
years, commencing at the present epoch and an orbital configuration
output every 100 simulation years. We used a simulation time
resolution of 1.0~day to provide adequate sampling based on the orbit
of the inner planet, Mercury. An initial baseline simulation was
conducted without the additional planet, to ensure that the
eccentricity variations of the Solar System planets were properly
reconstructed in agreement with other Solar System dynamical models,
such as those by \citet{laskar1988b}. The baseline simulation provided
an important point of comparison from which to evaluate the potential
divergence of the main suite of simulations with the additional
terrestrial planet. The divergence of the simulation results from the
baseline model were primarily tracked via the eccentricity evolution
of the Solar System planets, where planets are considered removed if
they are either ejected from the system or lost to the potential well
of the host star.


\subsection{Examples of Instability}
\label{examples}

As mentioned in Section~\ref{sim}, several thousand simulations were
conducted, producing a vast variety of dynamical outcomes for the
Solar System planets. The inner Solar System planets are particularly
vulnerable to the addition of the super-Earth planet, resulting in
numerous regions of substantial system instability. The broad region
of 2--4~AU contains many locations of MMR with the inner planets that
further amplify the chaotic evolution of the inner Solar System. There
are also important MMR locations with the outer planets within the
2--4~AU region, with potential significant consequences for the ice
giants. Here we provide several specific examples of simulation
results to highlight the range of possible dynamical outcomes.

\begin{figure*}
  \begin{center}
    \includegraphics[angle=270,width=16.0cm]{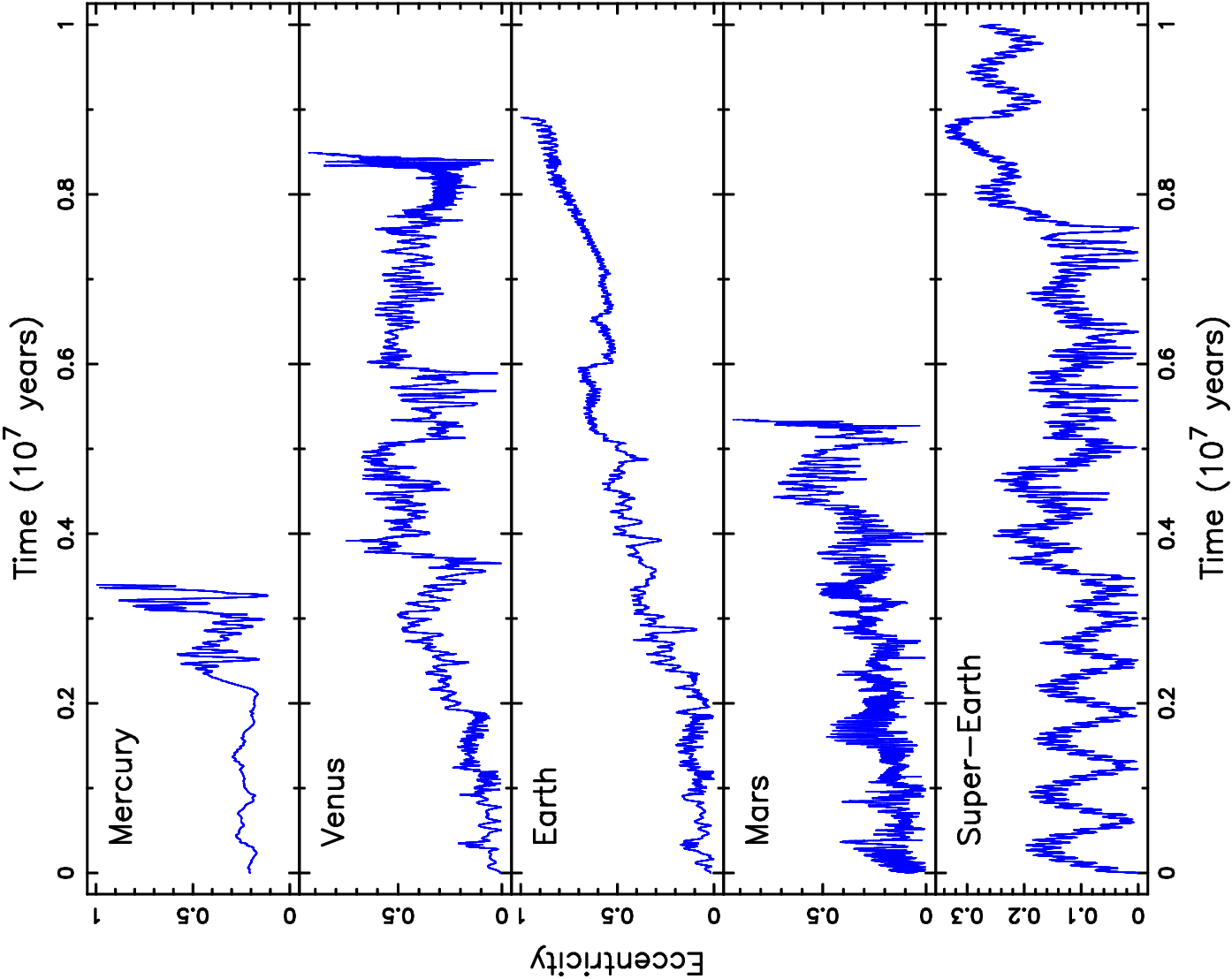}
  \end{center}
  \caption{Eccentricity evolution of the Solar System terrestrial
    planets (top four panels) for a $10^7$ year simulation, where the
    additional planet (bottom panel) has a mass and semi-major axis of
    7.0~$M_\oplus$ and 2.00~AU, respectively.}
  \label{fig:inner1a}
\end{figure*}

\begin{figure*}
  \begin{center}
    \includegraphics[angle=270,width=16.0cm]{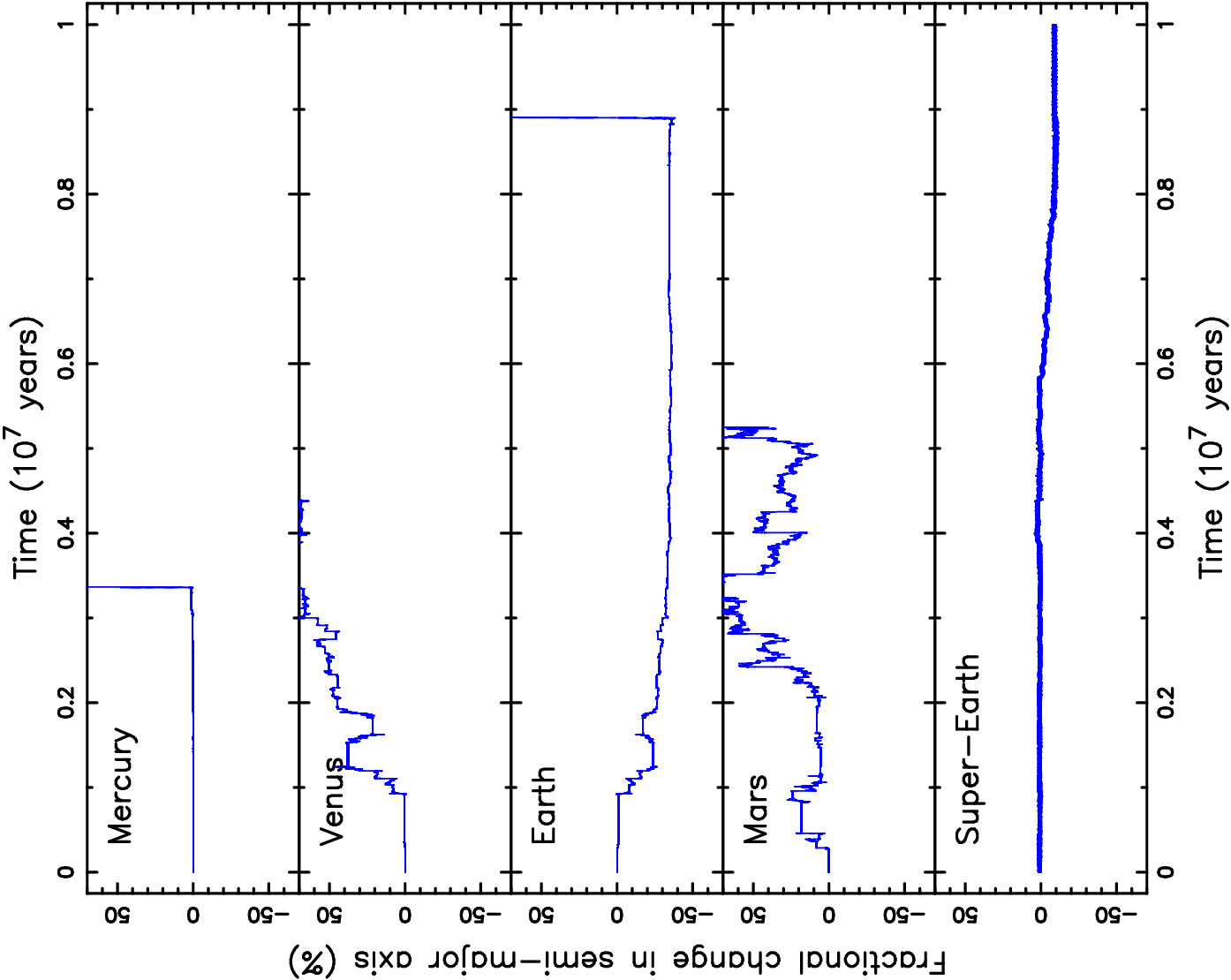}
  \end{center}
  \caption{Semi-major axis evolution, expressed as a fractional change
    from the initial value, of the Solar System terrestrial planets
    (top four panels) for a $10^7$ year simulation, where the
    additional planet (bottom panel) has a mass and semi-major axis of
    7.0~$M_\oplus$ and 2.00~AU, respectively.}
  \label{fig:inner1b}
\end{figure*}

Shown in Figure~\ref{fig:inner1a} is an example of the eccentricity
evolution of the inner planets for the case of an inserted planet with
mass and semi-major axis of 7.0~$M_\oplus$ and 2.00~AU,
respectively. The eccentricity evolution for the inserted planet is
shown in the bottom panel. In addition, we provide the evolution of
the semi-major axis for all planets in the panels of
Figure~\ref{fig:inner1b}, shown as the fractional change in the
semi-major axis from the initial value. For this example, the orbits
of all four inner planets become sufficiently unstable such that they
are removed from the system before the conclusion of the $10^7$ year
simulation. The semi-major axis of 2~AU places the additional planet
in a 2:3 MMR with Mars, leading to a rapid deterioration of the
Martian orbit, culminating with the ejection of Mars approximately
halfway through the simulation. Mercury is also ejected relatively
early in response to interactions with Venus and Earth, whose
eccentricities gradually increase and deposit angular momentum into
the Mercury orbit. Figure~\ref{fig:inner1a} and
Figure~\ref{fig:inner1b} show the gradual increase in eccentricities
for Venus and Earth, along with the increasing semi-major axis of
Venus and decreasing semi-major axis of Earth, leading to catastrophic
close encounters. Consequently, both Venus and Earth are removed from
the system during the time period of 8--9~Myrs after the commencement
of the simulation. For the super-Earth itself, an examination of the
periastron longitude relative to those of Jupiter and Saturn reveal a
weak connection of the eccentricity evolution to the $\nu_6$ secular
resonance with Saturn, and is largely affected by the presence of
Jupiter.

\begin{figure*}
  \begin{center}
    \includegraphics[angle=270,width=16.0cm]{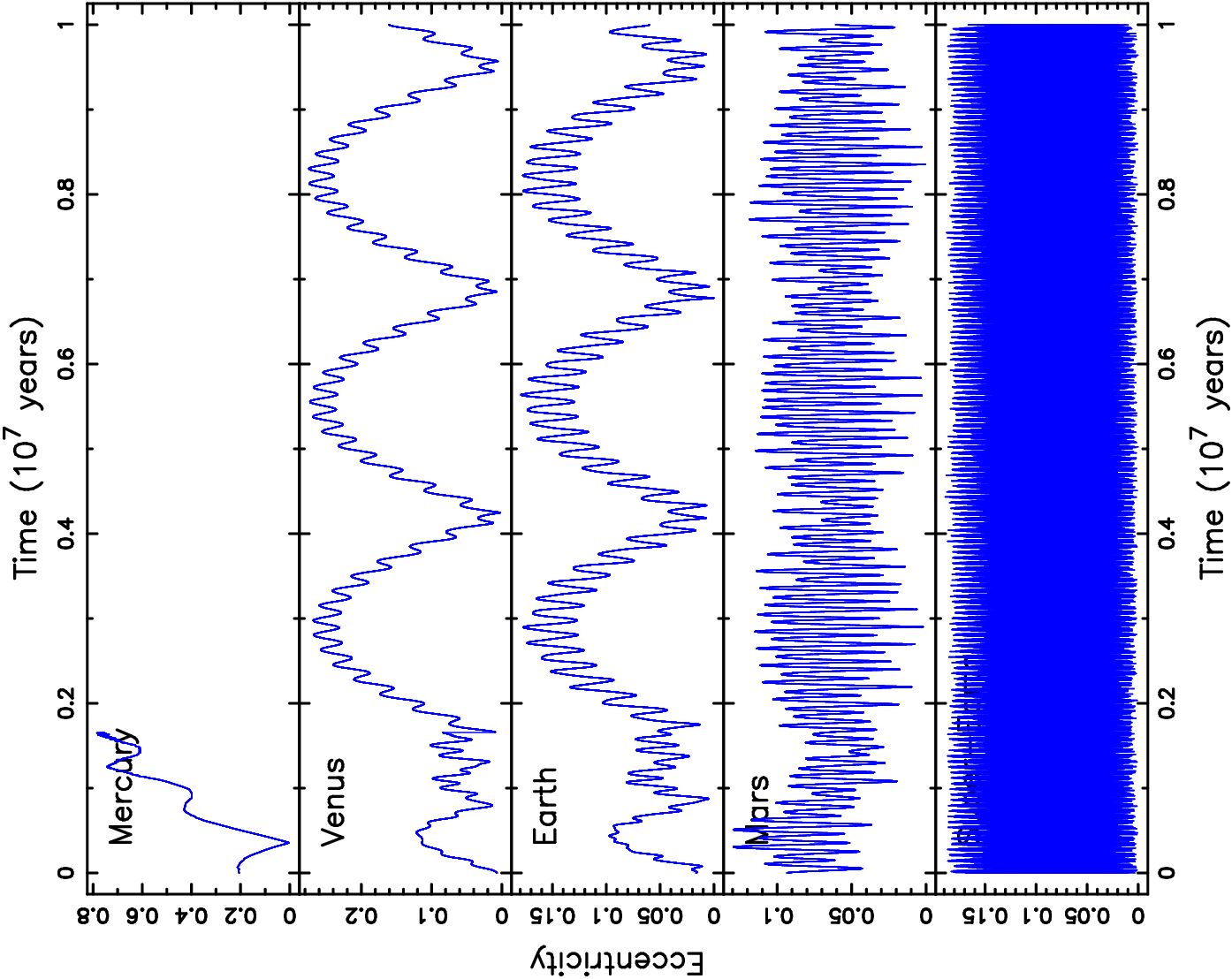}
  \end{center}
  \caption{Eccentricity evolution of the Solar System terrestrial
    planets (top four panels) for a $10^7$ year simulation, where the
    additional planet (bottom panel) has a mass and semi-major axis of
    8.0~$M_\oplus$ and 3.70~AU, respectively.}
  \label{fig:inner2a}
\end{figure*}

\begin{figure*}
  \begin{center}
    \includegraphics[angle=270,width=16.0cm]{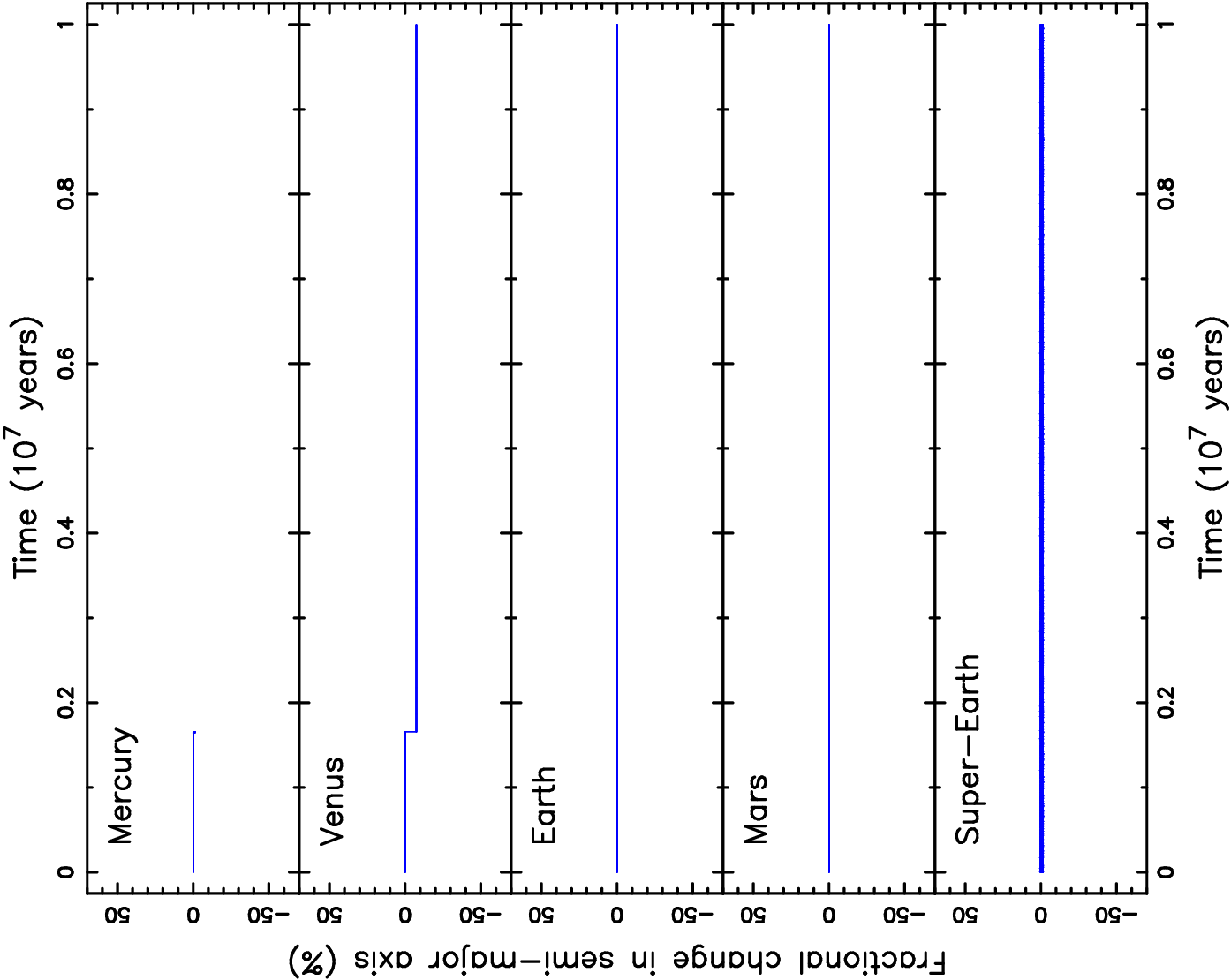}
  \end{center}
  \caption{Semi-major axis evolution, expressed as a fractional change
    from the initial value, of the Solar System terrestrial planets
    (top four panels) for a $10^7$ year simulation, where the
    additional planet (bottom panel) has a mass and semi-major axis of
    8.0~$M_\oplus$ and 3.70~AU, respectively.}
  \label{fig:inner2b}
\end{figure*}

A second example for the inner Solar System planets is shown in both
Figure~\ref{fig:inner2a} and Figure~\ref{fig:inner2b}, for the case of
an inserted planet with mass and semi-major axis of 8.0~$M_\oplus$ and
3.70~AU, respectively. In this case, the orbit of Mars is relatively
unaffected by the presence of the additional planet, whose
eccentricity undergoes high frequency oscillations due to interactions
with the outer planets. However, the $\nu_5$ secular resonance with
Jupiter plays a role in the eccentricity excitations of Venus and
Earth. In particular, the intial slight increase in the eccentricities
of Venus and Earth, combined with the $\nu_5$ resonance, is sufficient
to perturb the orbit of Mercury, resulting in the rapid removal of
Mercury from the system. The catastrophic loss of Mercury, causes a
subsequent injection of that angular momentum into the orbits of Venus
and Earth. This results in a substantial periodic evolution of their
orbits, with both high and low frequency variations in their
eccentricities. Figure~\ref{fig:inner2b} shows that, although the
semi-major axis of Earth and Mars are laregly unaltered, the
semi-major axis of Venus experiences a slight reduction.

\begin{figure*}
  \begin{center}
    \includegraphics[angle=270,width=16.0cm]{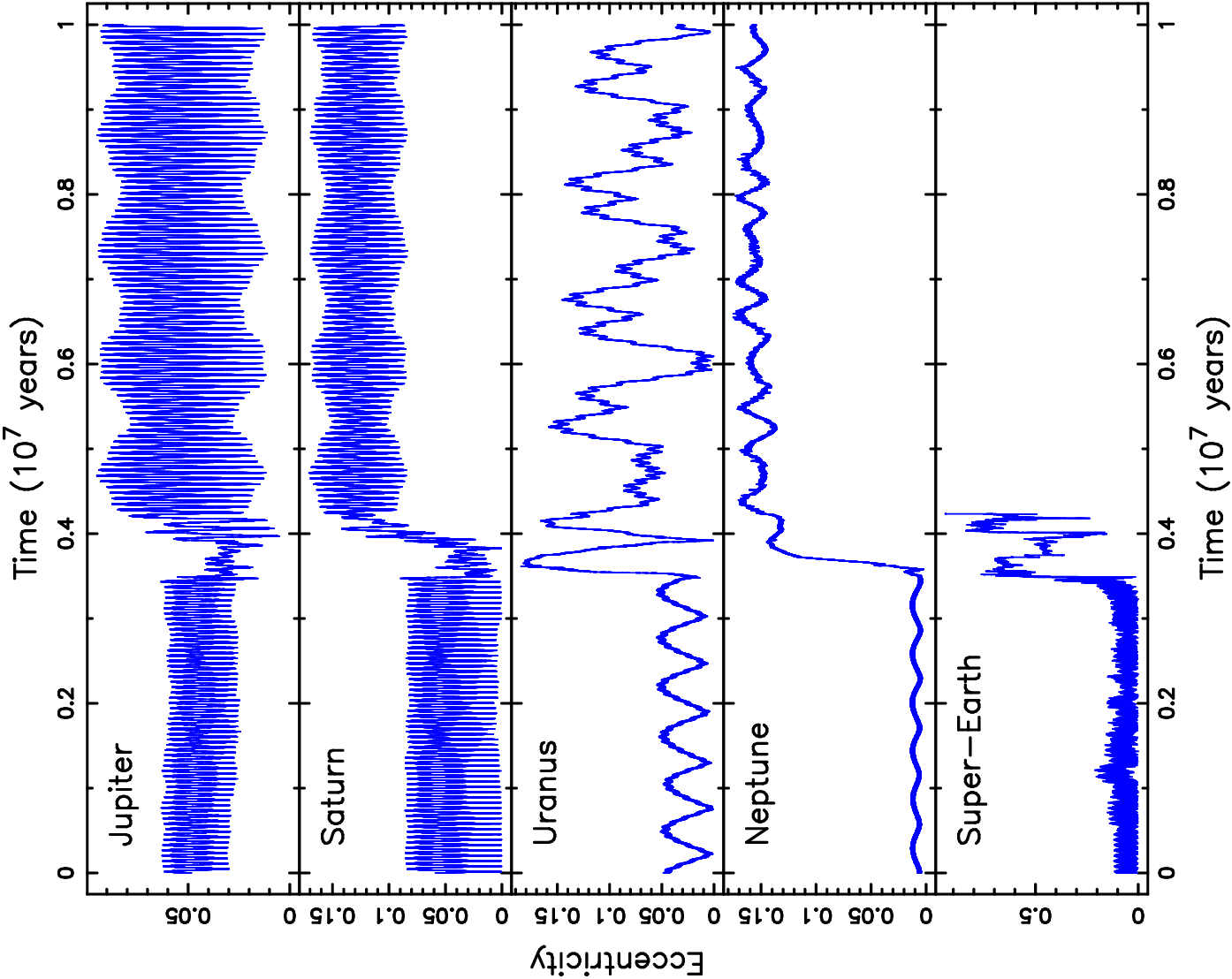}
  \end{center}
  \caption{Eccentricity evolution of the Solar System outer planets
    (top four panels) for a $10^7$ year simulation, where the
    additional planet (bottom panel) has a mass and semi-major axis of
    7.0~$M_\oplus$ and 3.79~AU, respectively.}
  \label{fig:outer1a}
\end{figure*}

\begin{figure*}
  \begin{center}
    \includegraphics[angle=270,width=16.0cm]{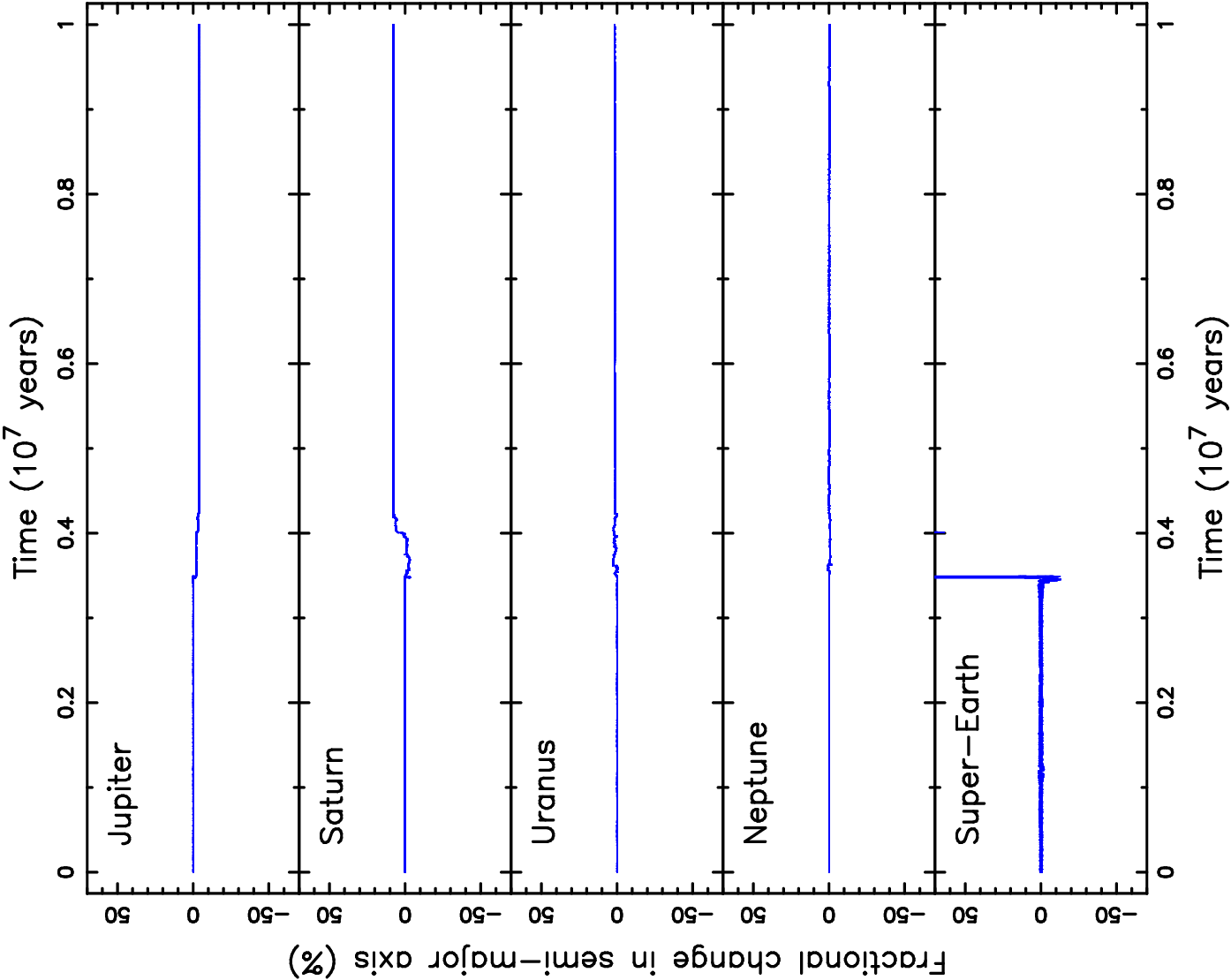}
  \end{center}
  \caption{Semi-major axis evolution, expressed as a fractional change
    from the initial value, of the Solar System outer planets
    (top four panels) for a $10^7$ year simulation, where the
    additional planet (bottom panel) has a mass and semi-major axis of
    7.0~$M_\oplus$ and 3.79~AU, respectively.}
  \label{fig:outer1b}
\end{figure*}

The effect of the additional planet on the orbital evolution of the
outer planets tends to be less severe than for the inner planets, but
there are still several cases of significant orbital
perturbations. The example shown in Figure~\ref{fig:outer1a} and
Figure~\ref{fig:outer1b} reveals the orbital evolution of the outer
planets for the case where the additional planet has a mass and
semi-major axis of 7.0~$M_\oplus$ and 3.79~AU, respectively. A
significant aspect of this orbital configuration is that the
additional planet lies at the 8:5 MMR with Jupiter, creating further
opportunity for orbital instability within the system. Indeed, the
orbits of the planets are largely unaffected for the first several
million years. After $\sim$3.5~Myrs, a dramatic change occurs within
the system, whereby the orbit of the super-Earth planet is
substantially altered, with high eccentricity (0.5--0.7) and with
semi-major axis values in the range 10--30~AU, as indicated in the
bottom panels of Figure~\ref{fig:outer1a} and
Figure~\ref{fig:outer1b}. Figure~\ref{fig:outer1b} also shows that
this change in the super-Earth orbit corresponds with a slight
increase in the semi-major axis of Saturn. This period of instability
persists for the next several hundred thousand years until the
super-Earth is ejected from the system after a total integration time
of $\sim$4.2~Myrs. The loss of the super-Earth planet, and subsequent
transfer of angular momentum, has a marginal effect on the orbit of
Jupiter, but a substantial effect on the eccentricities of Saturn,
Uranus, and Neptune. In particular, the orbit of Uranus remains in a
quasi-perturbed state, and the third and fourth panels of
Figure~\ref{fig:outer1a} show that Uranus and Neptune thereafter
exchange angular momentum with a significantly higher amplitude than
their previous baseline interactions.

\begin{figure*}
  \begin{center}
    \includegraphics[angle=270,width=16.0cm]{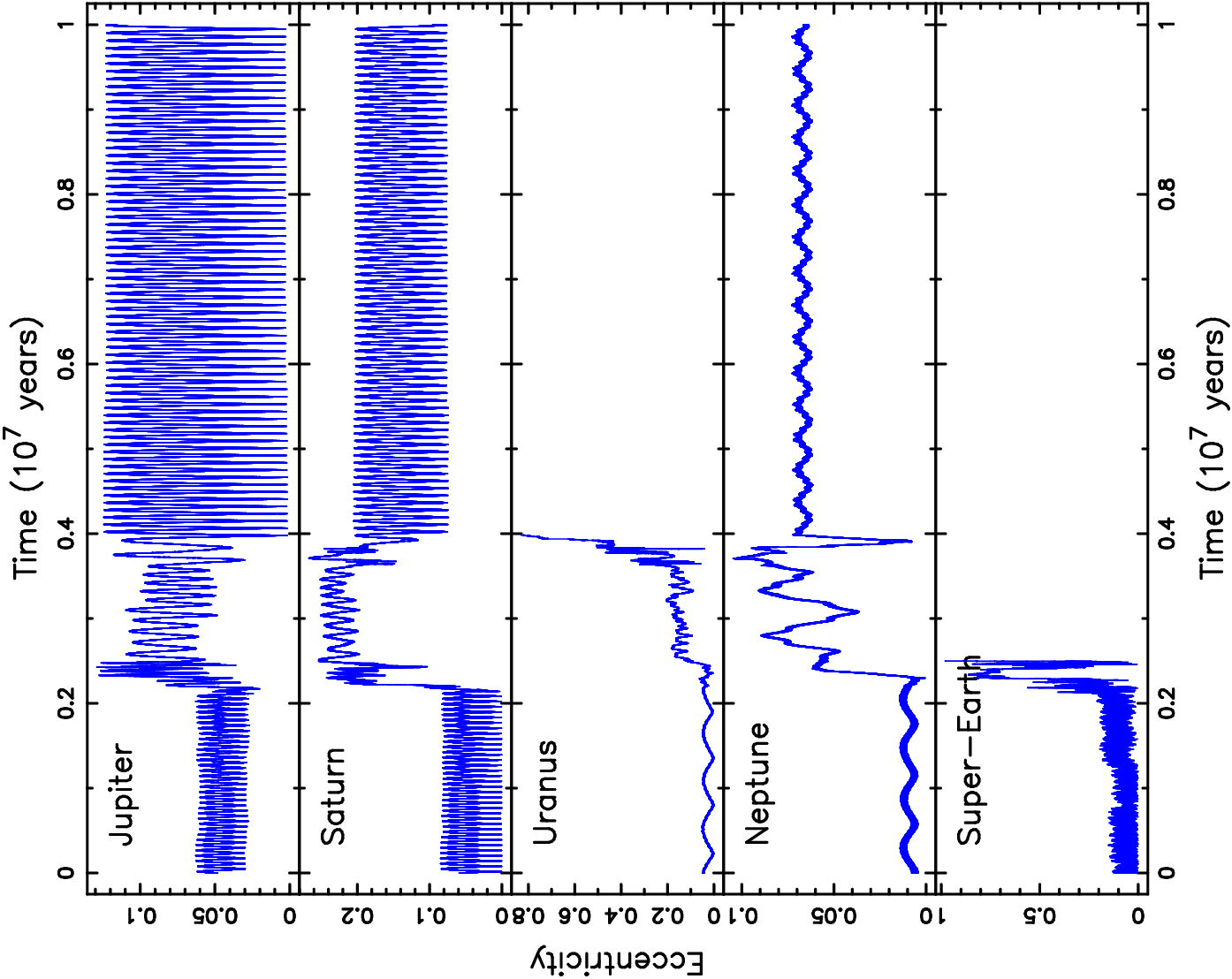}
  \end{center}
  \caption{Eccentricity evolution of the Solar System outer planets
    (top four panels) for a $10^7$ year simulation, where the
    additional planet (bottom panel) has a mass and semi-major axis of
    7.0~$M_\oplus$ and 3.80~AU, respectively.}
  \label{fig:outer2a}
\end{figure*}

\begin{figure*}
  \begin{center}
    \includegraphics[angle=270,width=16.0cm]{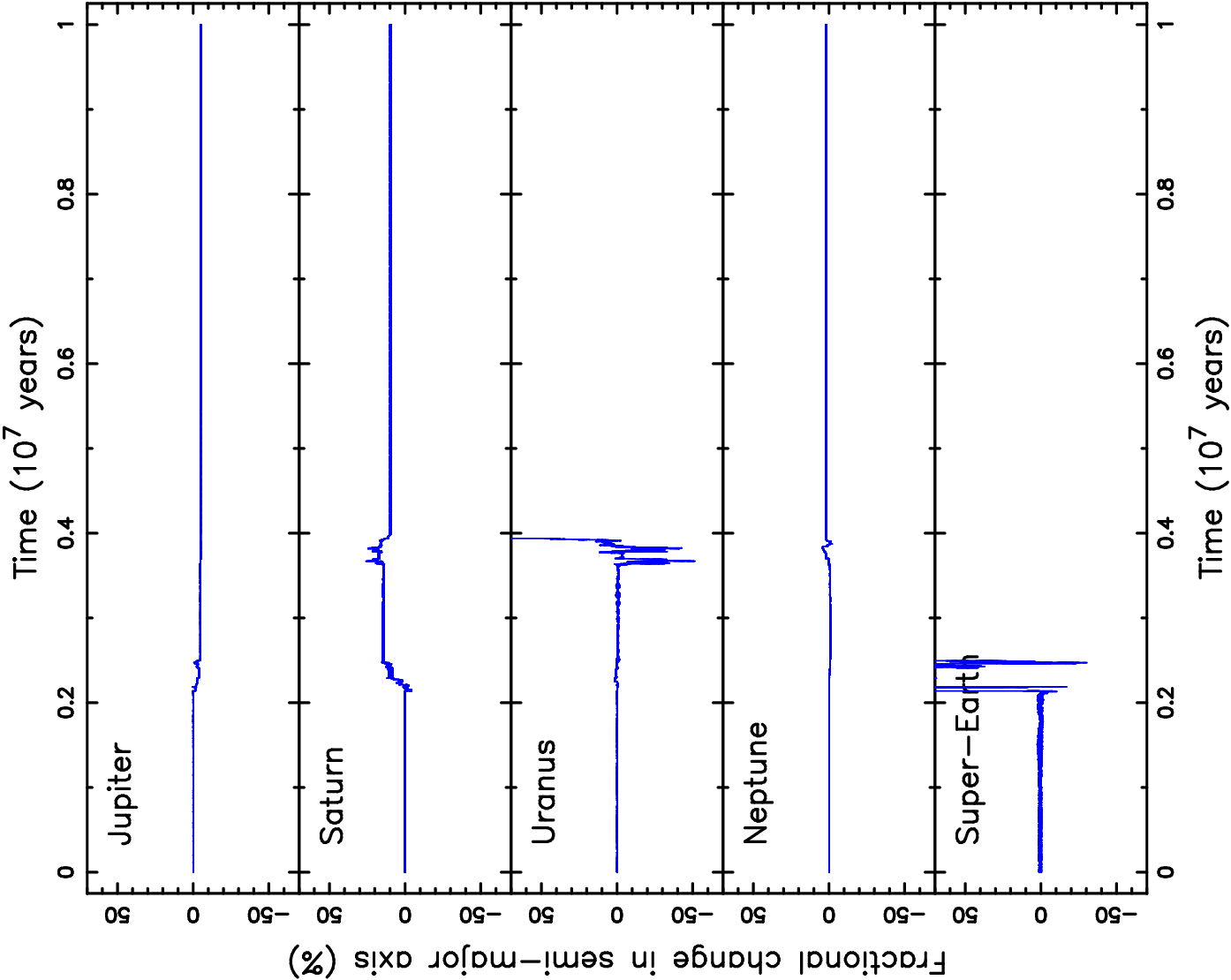}
  \end{center}
  \caption{Semi-major axis evolution, expressed as a fractional change
    from the initial value, of the Solar System outer planets
    (top four panels) for a $10^7$ year simulation, where the
    additional planet (bottom panel) has a mass and semi-major axis of
    7.0~$M_\oplus$ and 3.80~AU, respectively.}
  \label{fig:outer2b}
\end{figure*}

A final example of the outer planet eccentricity evolution is shown in
Figure~\ref{fig:outer2a} and Figure~\ref{fig:outer2b}. In this case,
the additional planet has a mass and semi-major axis of 7.0~$M_\oplus$
and 3.80~AU, respectively, so the overall system architecture is only
slightly different to the case shown in Figure~\ref{fig:outer1a} and
Figure~\ref{fig:outer1b}. Indeed, the super-Earth is similarly
influenced by the 8:5 MMR with Jupiter and the 4:1 MMR with Saturn,
leading to significant planet-planet interactions after
$\sim$2.5~Myrs. These interactions are especially manifest in the
increased eccentricity of Jupiter and Saturn, the latter of which also
increases in semi-major axis, and the ejection of the super-Earth from
the system entirely. These interations also perturb the orbit of
Neptune, which moves through a period of chaotic changes in
eccentrcity up until $\sim$4~Myrs. The biggest effect of the
interactions though is the loss of Uranus after $\sim$4~Myrs,
illustrating how the presence of a super-Earth within the Solar System
can greatly alter the architecture of the outer planets. The
substantial difference between the cases shown in
Figure~\ref{fig:outer1a} and Figure~\ref{fig:outer2a} demonstrates the
sensitivity of the dynamical outcome to the initial conditions,
particularly near locations of MMR. It is also worth noting that, in
cases such as these that produce significant interactions in the outer
Solar System, the orbits of the inner planets generally become
unstable also. For example, in the case described here and represented
in Figure~\ref{fig:outer2a} and Figure~\ref{fig:outer2b}, Mars is
ejected from the Solar System after $\sim$6~Myrs.

\subsection{System-Wide Results}
\label{results}

As described in Section~\ref{sim}, several thousand simulations with
the inclusion of a super-Earth planet were executed, in addition to a
baseline simulation for the unmodified Solar System architecture. The
divergence from the baseline scenario was evaluated by dividing the
eccentricity range throughout the simulation for each planet with the
eccentricity range from the baseline case. This produced a ratio of
eccentricity ranges that are generally $\geq 1$, where unity implies
that there is relatively little change in the orbital evolution of
that planet for a particular simulation. Cases where the ratio of
eccentricity ranges is $< 1$ can occur in relatively rare situations
where the damping of eccentricity oscillations exceeds the
perturbative effects of the other planets.

\begin{figure*}
  \begin{center}
    \includegraphics[angle=270,width=16.0cm]{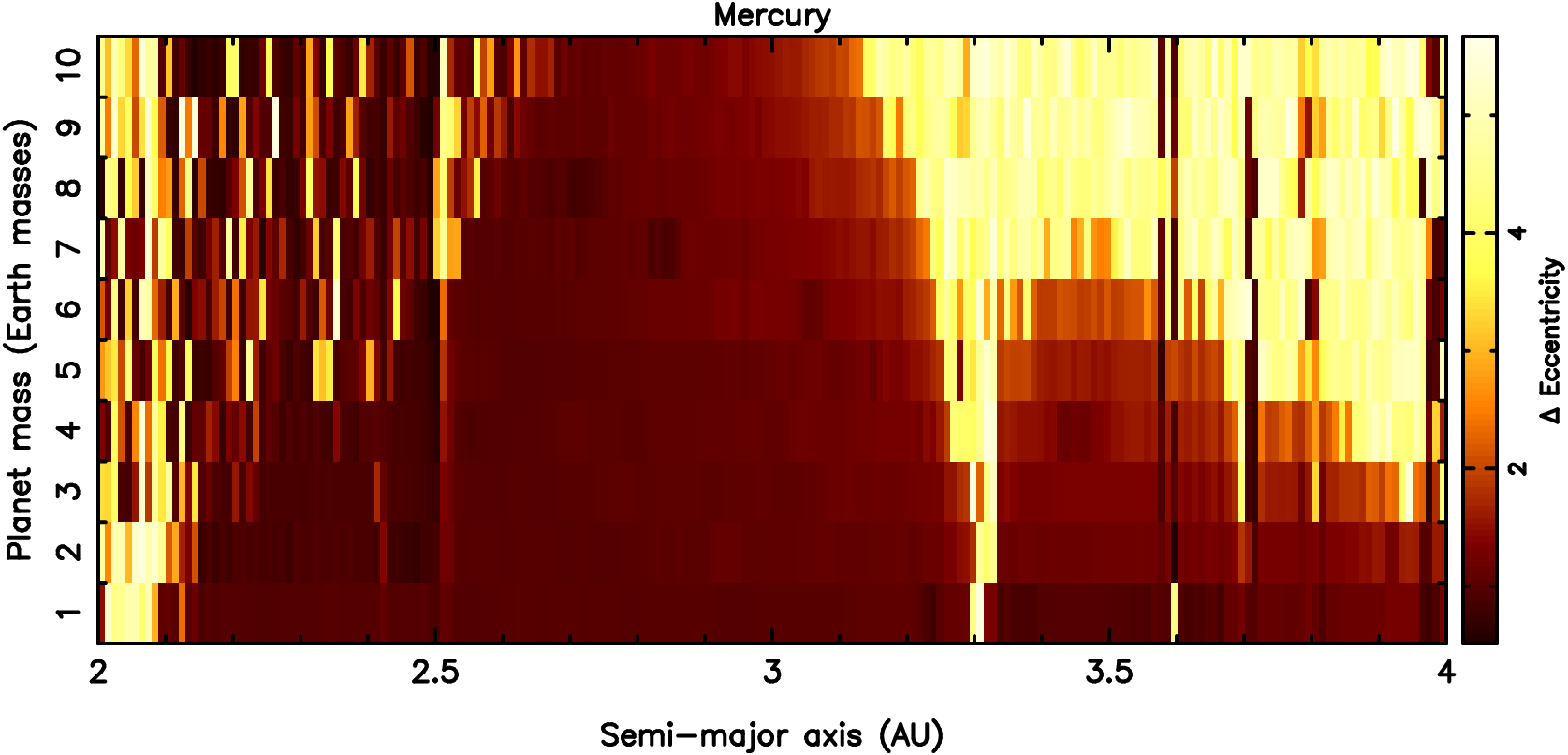} \\
    \vspace{0.5cm}
    \includegraphics[angle=270,width=16.0cm]{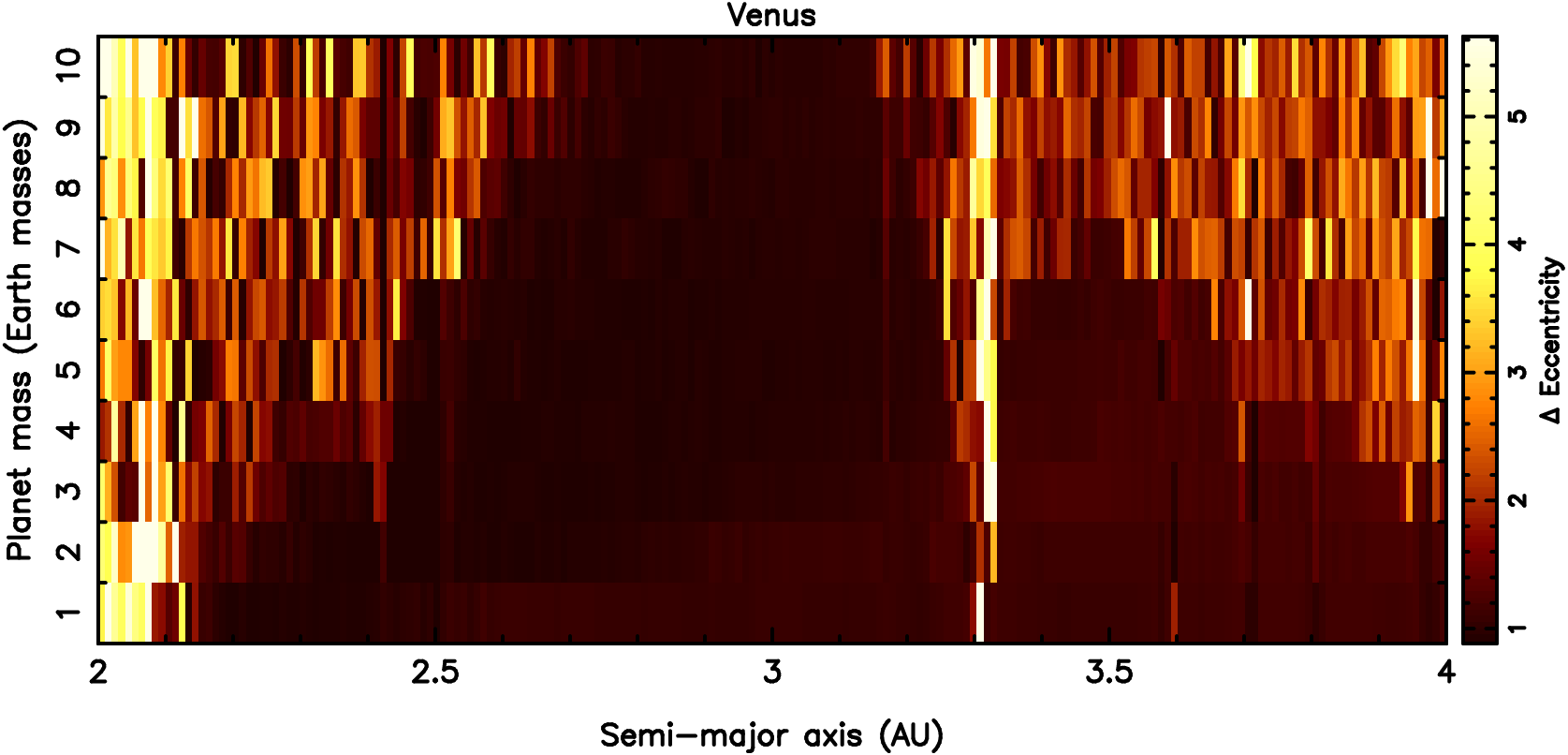}
    \end{center}
  \caption{Intensity plot showing the ratio of eccentricity variation
    amplitude to those from the baseline Solar System model, as a
    function of added planet mass and semi-major axis. Results are
    shown for Mercury (top panel) and Venus (bottom panel).}
  \label{fig:p12}
\end{figure*}

\begin{figure*}
  \begin{center}
    \includegraphics[angle=270,width=16.0cm]{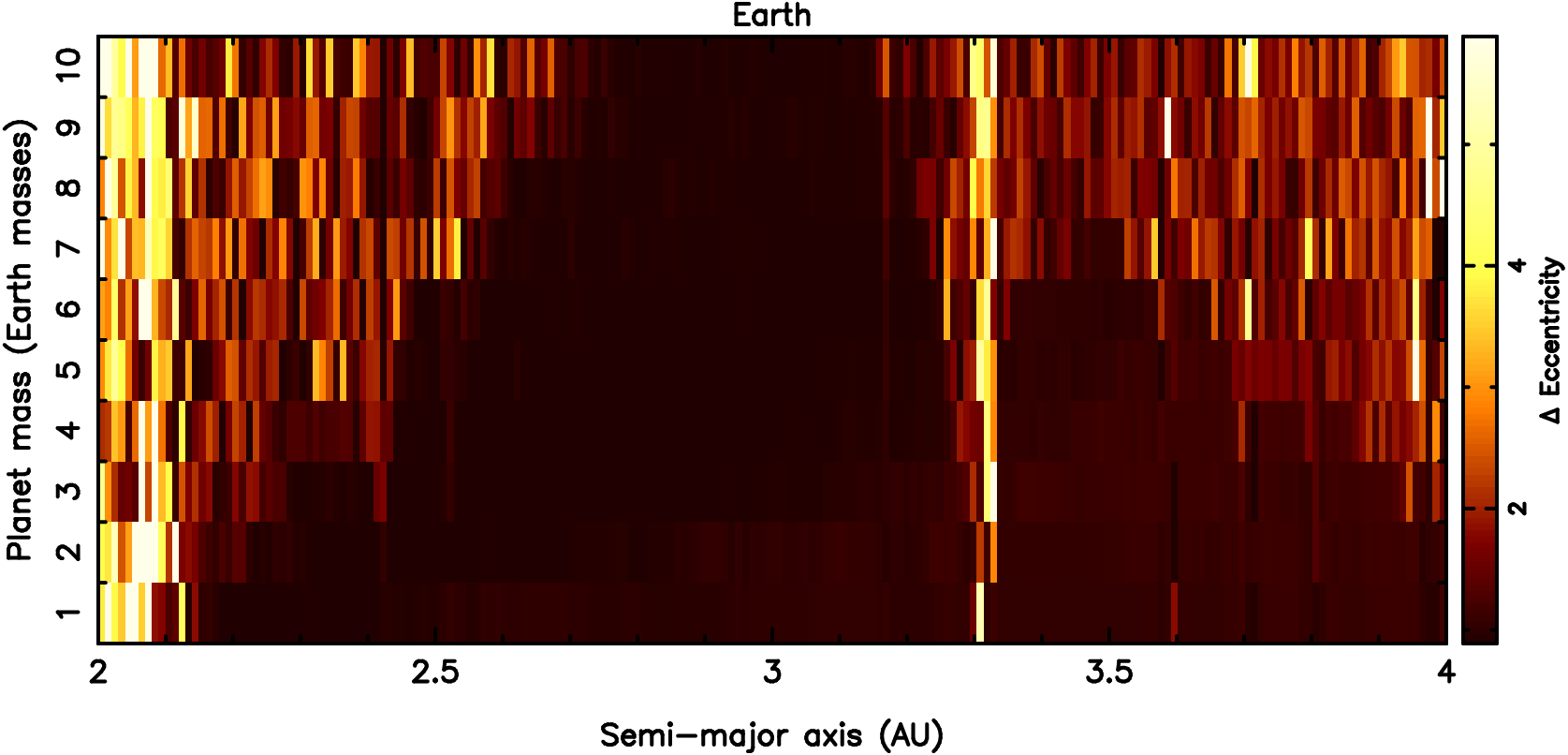} \\
    \vspace{0.5cm}
    \includegraphics[angle=270,width=16.0cm]{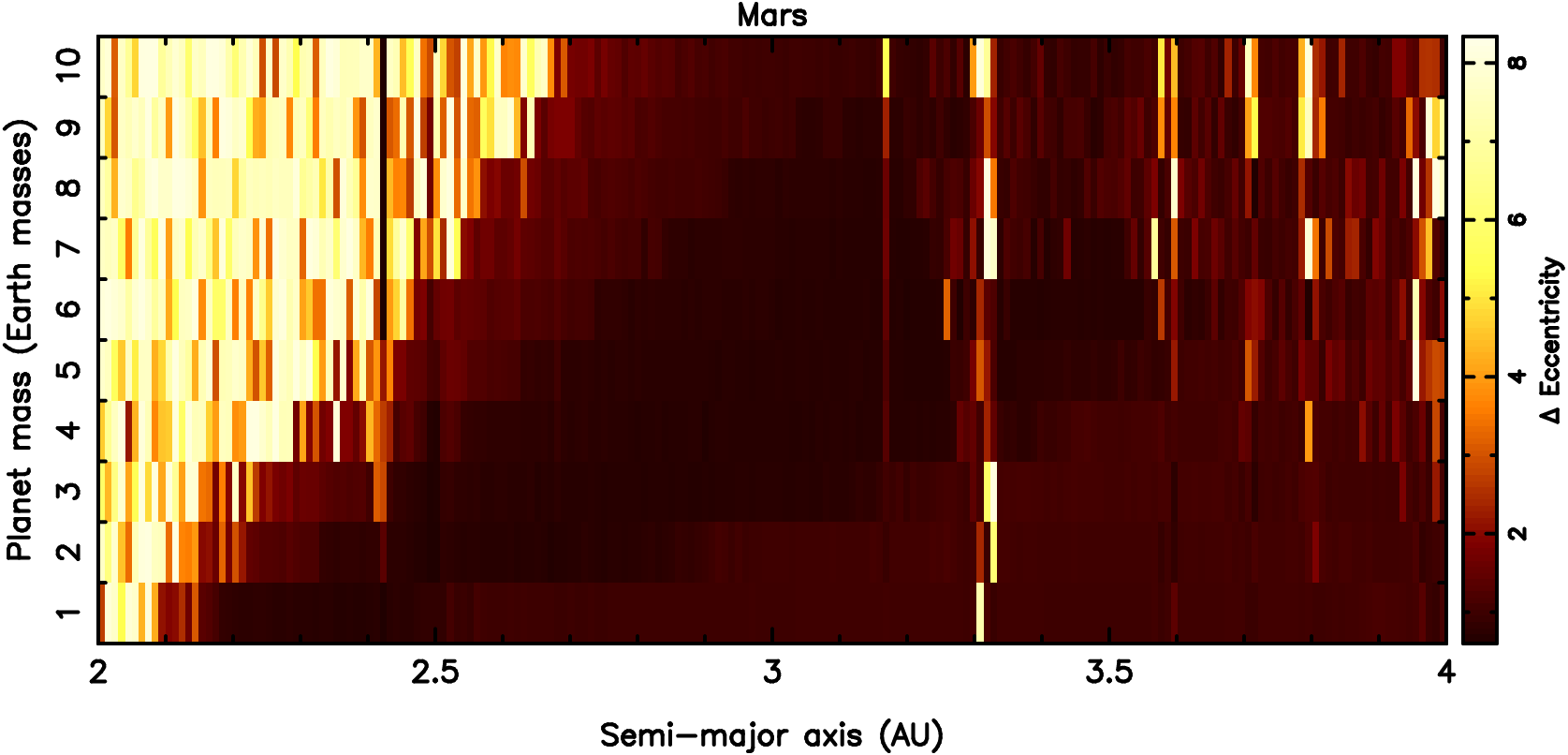}
    \end{center}
  \caption{Intensity plot showing the ratio of eccentricity variation
    amplitude to those from the baseline Solar System model, as a
    function of added planet mass and semi-major axis. Results are
    shown for Earth (top panel) and Mars (bottom panel).}
  \label{fig:p34}
\end{figure*}

\begin{figure*}
  \begin{center}
    \includegraphics[angle=270,width=16.0cm]{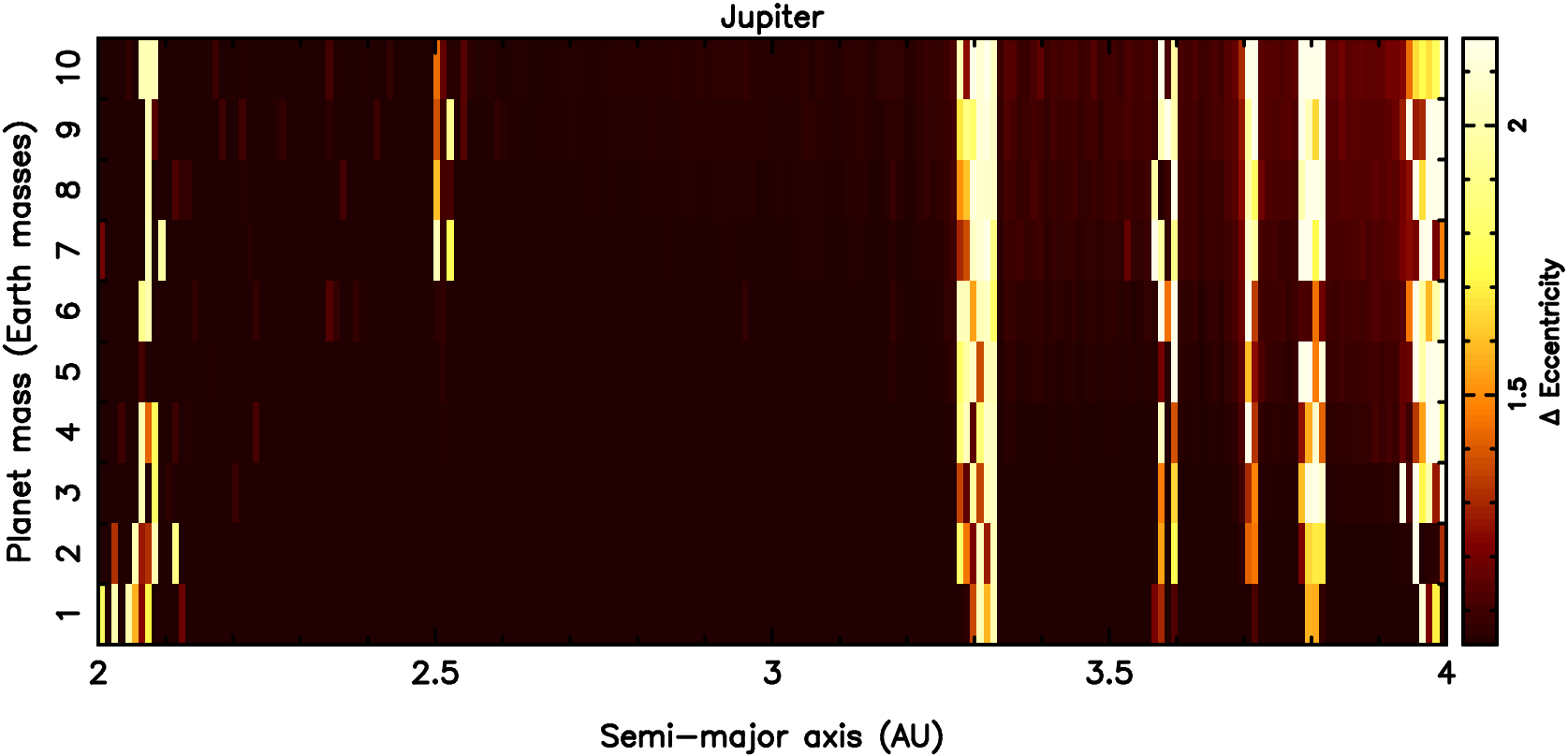} \\
    \vspace{0.5cm}
    \includegraphics[angle=270,width=16.0cm]{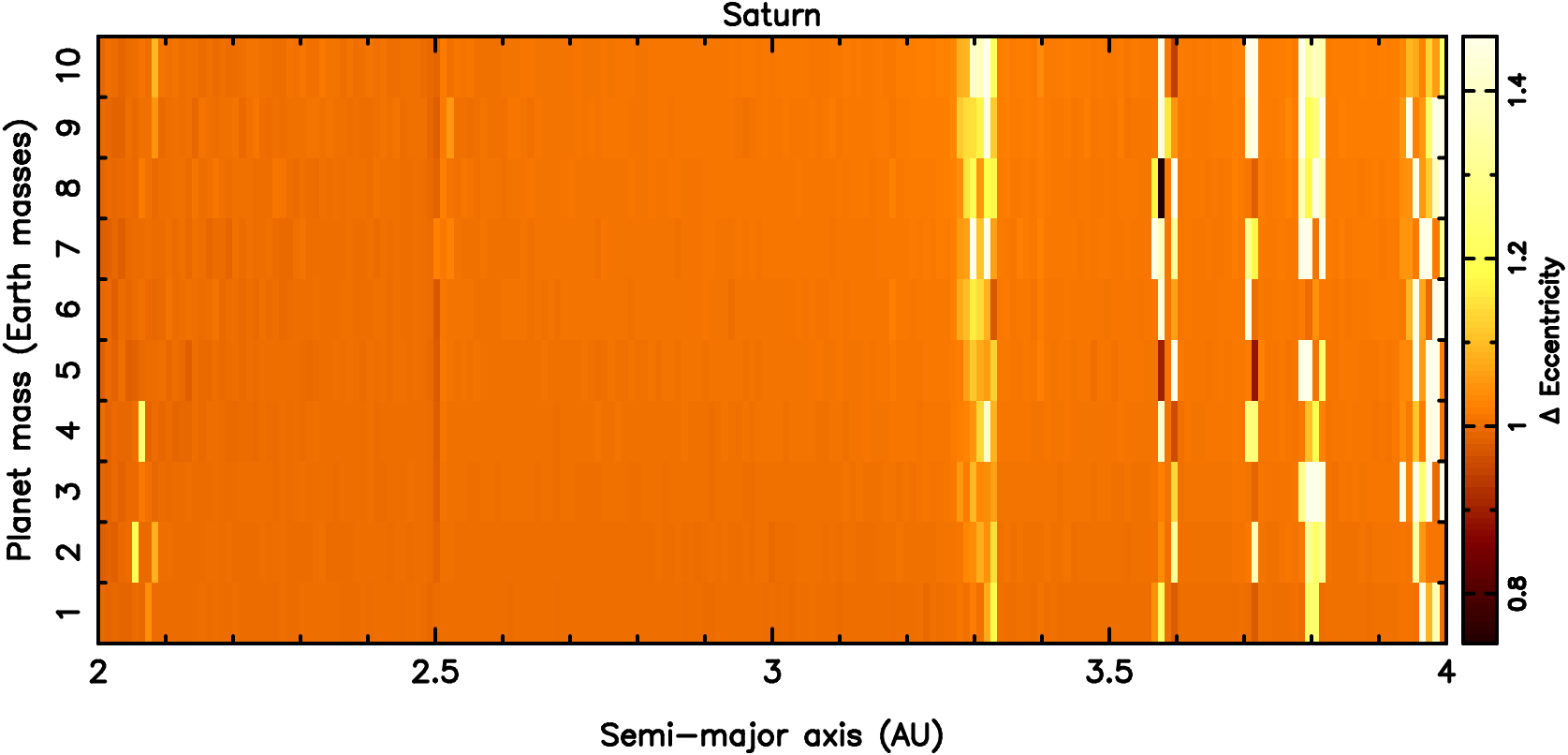}
    \end{center}
  \caption{Intensity plot showing the ratio of eccentricity variation
    amplitude to those from the baseline Solar System model, as a
    function of added planet mass and semi-major axis. Results are
    shown for Jupiter (top panel) and Saturn (bottom panel).}
  \label{fig:p56}
\end{figure*}

\begin{figure*}
  \begin{center}
    \includegraphics[angle=270,width=16.0cm]{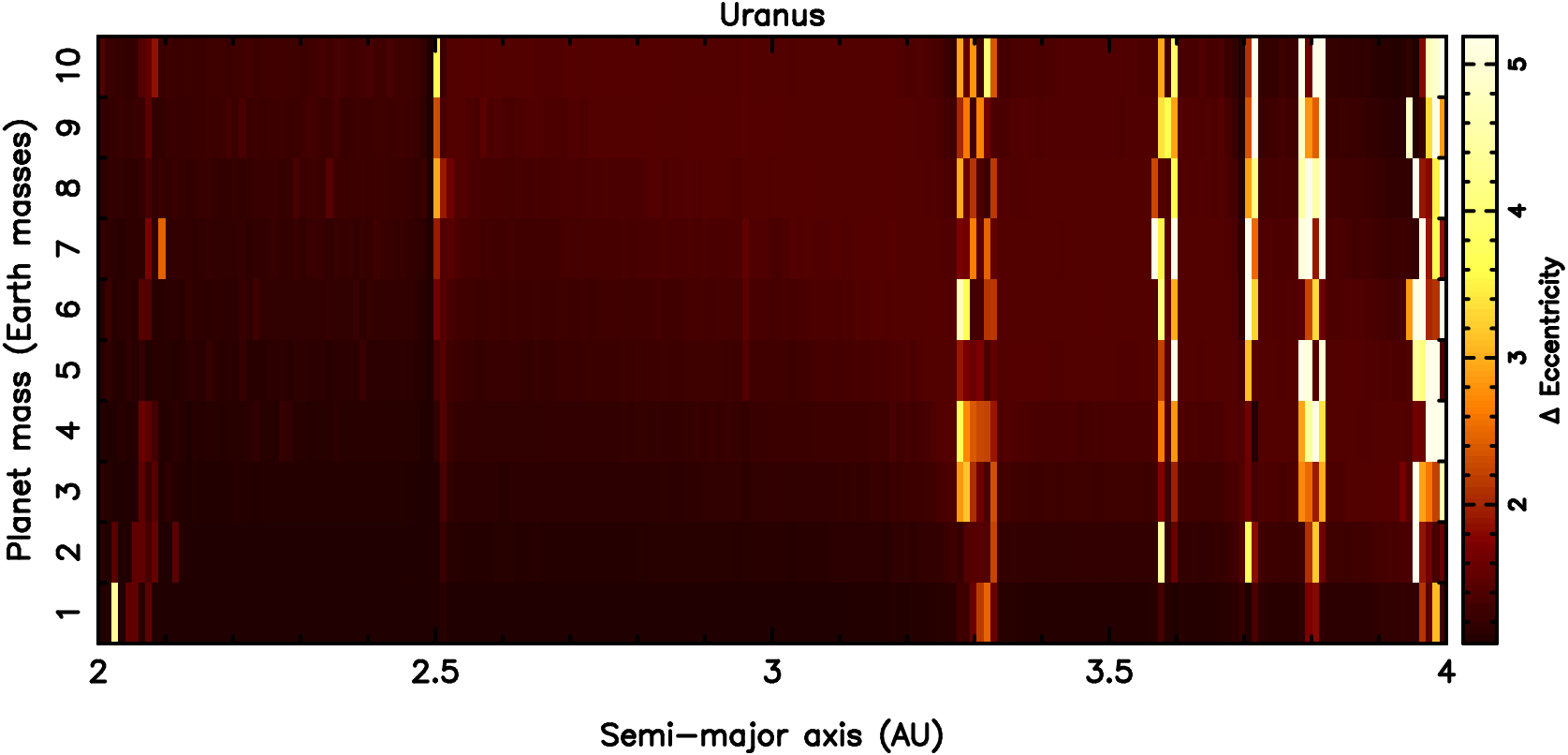} \\
    \vspace{0.5cm}
    \includegraphics[angle=270,width=16.0cm]{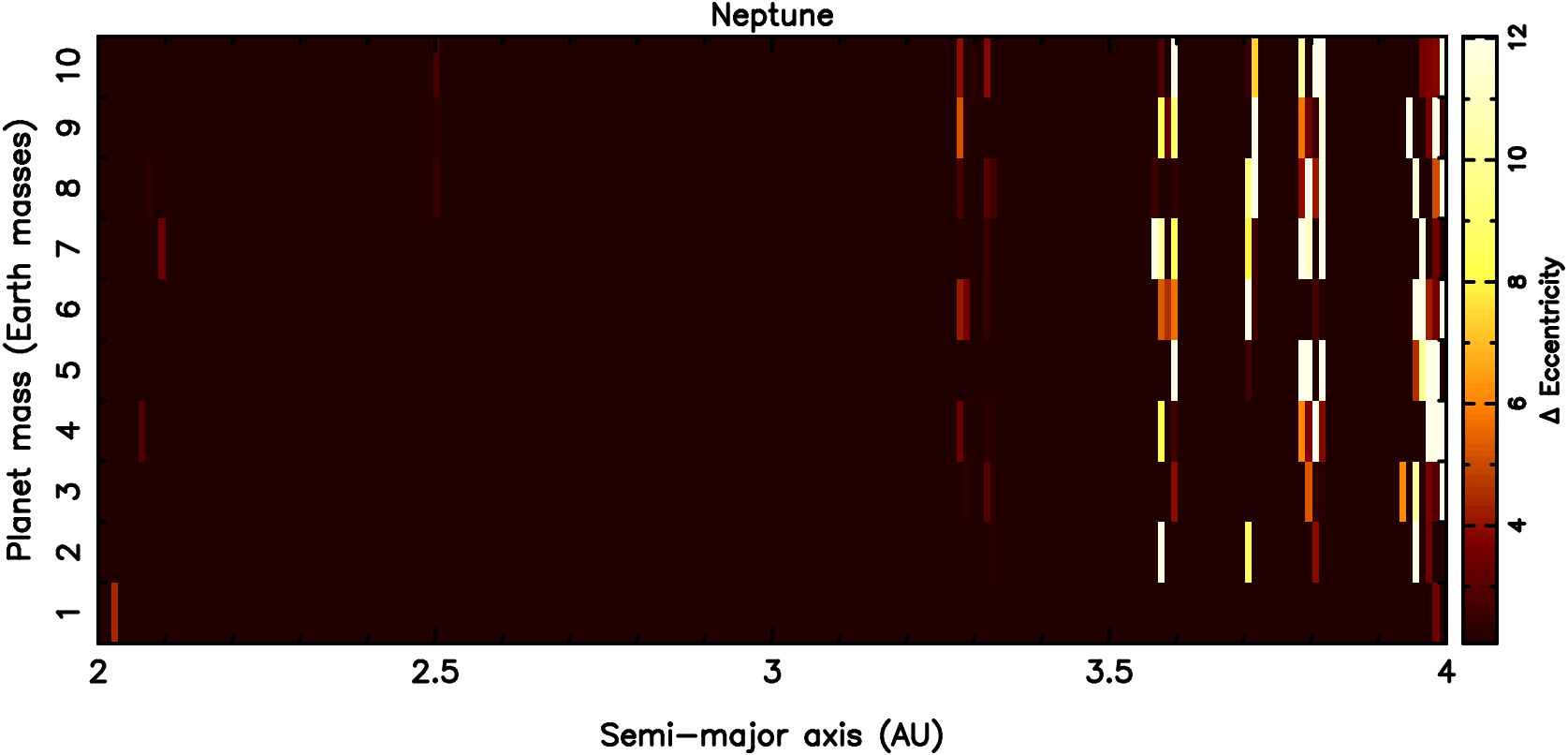}
    \end{center}
  \caption{Intensity plot showing the ratio of eccentricity variation
    amplitude to those from the baseline Solar System model, as a
    function of added planet mass and semi-major axis. Results are
    shown for Uranus (top panel) and Neptune (bottom panel).}
  \label{fig:p78}
\end{figure*}

The results of the above described calculations for all of the
simulations are represented as intensity plots, shown in Figures
\ref{fig:p12}, \ref{fig:p34}, \ref{fig:p56}, and \ref{fig:p78}. The
calculated eccentricity ranges relative to the baseline model are
labeled ``$\Delta$ Eccentricity'' on the color bars. Here, we note
some of the major features of these figures, and discuss their
implications in Section~\ref{con}. Figure~\ref{fig:p12} shows the
results for Mercury and Venus, and Figure~\ref{fig:p34} shows the
results for Earth and Mars. These figures demonstrate the sensitivity
of the inner planet orbits to the addition of the super-Earth at
locations in the range 2.0--2.7~AU, particularly for Mars. The orbit
of Mercury is also especially vulnerable to a super-Earth location in
the range 3.1--4.0~AU, resulting in significant eccentricity
excursions for Mercury. Venus and Earth appear to have very similar
orbital evolution responses to the addition of the super-Earth,
regardless of the planetary mass or semi-major axis location.

Figure~\ref{fig:p56} shows the results for Jupiter and Saturn, and
Figure~\ref{fig:p78} shows the results for Uranus and Neptune. The
features of the eccentricity divergences for the outer planets from
the baseline model primarily occur at locations of MMR. Indeed, a
location of instability common to both the inner and outer planets is
$\sim$3.3~AU which is the 2:1 MMR with Jupiter and a significant
location within the Kirkwood gaps (see discussion in
Section~\ref{why}). As expected, the variations in eccentricity are
relatively small for Jupiter and Saturn, who remain largely unaffected
by the presence of the super-Earth except at resonance
locations. However, as demonstrated by the several outer planet
examples in Section~\ref{examples}, the locations of MMR can have a
devastating effect on the orbits of Uranus and Neptune, including the
loss of one or both ice giants from the system.

\section{Dynamical Consequences}
\label{con}


\subsection{Implications for the Inner Planets}
\label{inner}

The results provided in Section~\ref{results}, and Figures
\ref{fig:p12} and \ref{fig:p34}, show that there are broad regions of
mass and semi-major axis for the additional planet that can have have
severe consequences for the inner Solar System planets. In general,
the inner planets retain their orbital integrity if the additional
planet is located within the range 2.7--3.1~AU, even for relatively
high masses. This semi-major axis region minimizes interactions with
the giant planets that would otherwise excite the orbital eccentricity
of the additional planet. However, the full range of semi-major axes
(2--4~AU) for the added super-Earth spans a period range of
1033--2922~days, in which there are numerous MMR locations with the
inner planets. For example, the semi-major axis range of 2.0-2.1~AU
encompasses the 1:3 MMR with Earth, though it is unclear that this MMR
plays a major contributing role to the destabilization of the other
inner planets. As shown in Figure~\ref{fig:p34}, the stability of Mars
is particularly sensitive to the super-Earth being located in the
semi-major axis range 2.0--2.7~AU. There are various MMR locations
throughout this region, including the 1:2 MMR with Mars at 2.42~AU,
and the 4:1 and 3:1 MMRs with Jupiter, located at 2.07~AU and 2.50~AU,
respectively.

Also of note is the evolution of the Mercury orbit into regions of
high eccentricity when the additional planet lies within the
semi-major axis range of 3.1--4.0~AU, occasionally resulting in the
ejection of Mercury from the system prior to the conclusion of the
simulation. Mercury does not lie near any significant locations of MMR
with the full range of semi-major axes for the additional
planet. However, there have been numerous studies regarding the
chaotic nature of Mercury's orbit
\citep{ward1976,correia2004,lithwick2011b,noyelles2014}, including its
sensitivity to the eccentricity and inclination of Venus'
orbit. Figures \ref{fig:p12} and \ref{fig:p34} show that the
eccentricities of the Venus and Earth orbits are impacted by the
3.1--4.0~AU semi-major axis range of the super-Earth as well, although
to a lesser extent than that of Mercury. Additionally, as noted in
Section~\ref{examples}, the orbits of Venus and Mercury are sensitive
to the $\nu_5$ resonance with Jupiter, and the combination of such
resonances may be the primary cause of the Venus orbital
changes. Consequently, these alterations to the Venus and Earth orbits
may be sufficient to dramatically contribute to the evolution of
Mercury's orbit, further demonstrating the complex dynamical
interactions that can occur between planets that are relatively widely
separated.


\subsection{Implications for the Outer Planets}
\label{outer}

As seen from the several simulation results examples described in
Section~\ref{examples}, the outer planets do not escape the
consequences of the super-Earth presence in the Solar
System. Section~\ref{results} mentions that Jupiter and Saturn are
largely unperturbed by the presence of the super-Earth. The exceptions
to this are locations of MMR, where Jupiter and Saturn experience mild
increases in eccentricity. Particularly strong locations of MMR are
those associated with Jupiter; specifically: $\sim$2.06~AU (4:1),
$\sim$2.5~AU (3:1), $\sim$3.3~AU (2:1), $\sim$3.58~AU (7:4),
$\sim$3.7~AU (5:3), and $\sim$3.97~AU (3:2). Many of these locations
are represented as Kirkwood gaps in the asteroid belt, as described in
Section~\ref{why}. Note that $\sim$3.79~AU corresponds to the 4:1 MMR
with Saturn, but there is little evidence that this MMR plays a
significant role in the orbital evolution of the super-Earth. Even at
locations of MMR with the super-Earth, the perturbation to the Jupiter
and Saturn eccentricities remain relatively small, and there are no
simulation outcomes in which either planet are ejected from the Solar
System.

However, the consequences of the super-Earth presence near the MMR
locations are far more severe for the ice giants, with frequent
excitation of their eccentricities, including rare occasions of
ejection of the ice giants from the Solar System. The elevation of the
ice giant eccentricities has two major sources. The first source is
via the angular momentum transfer through Jupiter and Saturn, whose
inflated eccentricities, though relatively small, deliver substantial
dynamical effects to the ice giants. The second source is the result
of a dramatic alteration of the super-Earth orbit, placing it in a
highly-eccentric chaotic orbit with a semi-major axis beyond the orbit
of Saturn. In such cases, the ice giants gravitationally interact
directly with the super-Earth, and develop their own divergent orbital
states. These interactions usually lead to the super-Earth being
ejected from the system, and it is unlikely that the ice giant orbits
will maintain long-term stability beyond the $10^7$~year
simulation. It is worth noting that these described sources of orbital
perturbation for the ice giants are layered on top of complicated
interactions between the super-Earth and the overall dynamics of the
system, potentially resulting in further sources of instability,
particularly related to secular frequencies with Jupiter.


\subsection{Implications for Exoplanets}
\label{exoplanets}

As the known exoplanet population has grown, the study of demographics
and their statistical relationship to planetary architectures has
become increasingly important. The occurrence rates of terrestrial
planets, including super-Earths, have been estimated from radial
velocity
\citep[e.g.,][]{howard2010b,bashi2020,fulton2021,rosenthal2021,sabotta2021}
and transit surveys
\citep[e.g.,][]{dressing2013,kopparapu2013b,kunimoto2020b,bryson2021}. The
results of these occurrence rate estimates vary enormously, depending
upon observational biases, analysis techniques, and the parameter
space being explored. In particular, most of these studies are heavily
biased toward short-period planets around M dwarfs, since the radial
velocity amplitude and transit probability/depths are maximized within
that regime. However, these analyses consistently indicate that
terrestrial planets are relatively common, with super-Earth planets
occurring in most of the detected planetary systems. The exception to
these planetary size distributions occur within a photo-evaporation
regime of close-in planets, where there is a relative dearth of
super-Earths and mini-Neptunes \citep{lopez2013,owen2013a,fulton2017}.

On the other hand, giant planets are relatively rare, particularly
beyond the snow line, with occurrence rates averaging around 10\% for
solar-type stars, depending on the exact stellar type and semi-major
axis range explored \citep{wittenmyer2020a,fulton2021}. Giant planets
are even scarcer around later-type stars
\citep{zechmeister2013,wittenmyer2016c}, lending further credence to
the hypothesis that the Solar System may not be representative of
planetary architectures. Indeed, the presence of giant planets can be
disruptive to the orbits of dynamically packed systems
\citep{kane2020b}, and \citet{mulders2021b} found that the occurrence
of hot super-Earths and cold giant planets are anticorrelated around M
dwarfs. These findings are important for the results presented in this
work, since a major reason for the regions of instability within the
Solar System architecture are due to the interaction of the
super-Earth with the outer giant planets.

An important feature of exoplanetary systems is their eccentricity
distribution, and the important role that eccentricity plays in the
evolution and stability of terrestrial planet climates
\citep{williams2002,dressing2010,kane2012e,linsenmeier2015,bolmont2016a,kane2017d,way2017a,kane2020e}. As
shown in Figure~\ref{fig:inner2a}, the presence of the super-Earth has
significant effects on the orbits for the inner terrestrial
planets. These interactions result in large amplitude oscillation of
Venus and Earth orbital eccentricities, creating Milankovitch cycles
that may potentially influence the long-term climate of these planets
\citep{deitrick2018a,deitrick2018b,horner2020a,kane2021a}. Given the
vast range of planetary system architrectures and terrestrial planet
climates, it is unclear how the high occurrence rate of super-Earths
may contribute to the overall evolution of terrestrial climates in
exoplanetary systems \citep{becker2017a}. The dependence of planetary
climates on orbital interactions with super-Earths will require
further atmospheric data and modeling to determine if the presence of
such planets (or lack thereof) can preferentially lead to
eccentricity-driven climate effects.


\section{Conclusions}
\label{conclusions}

Up until several decades ago, the architecture of the Solar System was
the template for general models of planetary system formation and
evolution. The discovery of exoplanets has revealed a diversity of
orbital configurations that has motivated detailed dynamical analyses
of the various processes that can explain the observed distribution of
system properties \citep{fang2012f,tremaine2012,tremaine2015}. The
presence of giant planets, and their possible migration, can have a
profound effect on the types of planets that are formed within a
system and their eventual stable orbits, as has been the case for the
early history of the Solar System. In particular, these giant planet
migration events may have influenced the terrestrial planet formation
processes in the inner Solar System and truncated the formation of the
most common type of planet thus far discovered: super-Earths.

The lack of a super-Earth within the Solar System may be considered an
example of a ``double-edged sword'' situation. The negative aspect of
this lack is that we have no local analog from which to draw in-situ
data that would aid enormously in constructing interior and
atmospheric models for the exoplanet super-Earths, the vast majority
of which were detected using indirect techniques. This work presents a
positive aspect of the lack of a local super-Earth, in demonstrating
the potential for orbital instability that such an additional
planetary mass may induce. Locations of the super-Earth near 3~AU
generally preserve system stability, but the full region explored of
2--4~AU is peppered with MMR locations with the other planets. For the
inner planets, Mercury's orbit is mostly rendered unstable when the
super-Earth is located in the region 3.1--4.0~AU, and Mars' orbit
becomes unstable when the super-Earth lies within 2.0--2.7~AU. For the
outer planets, significant MMR locations are mostly located beyond
3.0~AU, and can result in substantially chaotic orbits for the ice
giants.

There are several factors that were not incorporated into this
analysis that are worth exploring in future studies. Section~\ref{sim}
explains that the initial orbit for the additional planet was coplanar
with Earth. Mutual inclinations between planetary orbits plays a role
in overall system stability \citep{laskar1989,chambers1996},
particularly for large inclinations \citep{veras2004b,correia2011},
and may provide solutions to otherwise unstable architectures
\citep{kane2016d,masuda2020a}. It is therefore possible that there are
orbital inclinations for the super-Earth that may reveal further
locations of long-term stability, or else enhance unstable scenarios.
Furthermore, our simulations did not include tidal dissipation, which
would serve to circularize orbits and dampen the orbital energy gained
through angular momentum transfer
\citep{barnes2009b,laskar2012,bolmont2014}. However, inclusion of
tidal dissipation would not affect the results of this work given the
relatively wide spacing of the Solar System planets and the timescale
of our simulations. Moreover, tidal interaction with the
protoplanetary disk likely played a major role in producing the low
eccentricities of the Solar System giant planets
\citep{tsiganis2005b}.

As noted above, the lack of a super-Earth within the solar system is
primarily a product of the giant planet formation pathway that
occurred. However, as evidenced by the relatively low occurrence rate
of Jupiter analogs for solar-type stars (for example), such a
formation pathway is not guaranteed, and super-Earth formation can
occur in the presence of giant planets. Our results reveal the
dynamical fragility of our existing planetary configuration, allowing
a more detailed examination of this configuration within the broader
context of planetary system architectures. The primary application of
this work is thus via a comparison with the plethora of known
exoplanetary systems, especially given the apparent stability
sensitivity of the solar system to alternative architectures, such as
that described here. With numerous precision radial velocity surveys
moving forward
\citep[e.g.,][]{pepe2014a,fischer2016,stefansson2016,gupta2021} and
the era of space-based astrometry upon us \citep{brown2021}, the
Keplerian orbits of nearby planetary systems will be gradually
revealed and refined. The study of the orbits with these systems, both
from individual and statistical points of view, will demonstrate the
true consequences of sharing dynamical space with a super-Earth
planet.


\section*{Acknowledgements}

The author would like to thank Matthew Clement, Sean Raymond, and the
two anonymous referees for useful feedback on the manuscript. The
results reported herein benefited from collaborations and/or
information exchange within NASA's Nexus for Exoplanet System Science
(NExSS) research coordination network sponsored by NASA's Science
Mission Directorate.


\software{Mercury \citep{chambers1999}}



\begin{thebibliography}{}
\expandafter\ifx\csname natexlab\endcsname\relax\def\natexlab#1{#1}\fi
\providecommand{\url}[1]{\href{#1}{#1}}
\providecommand{\dodoi}[1]{doi:~\href{http://doi.org/#1}{\nolinkurl{#1}}}
\providecommand{\doeprint}[1]{\href{http://ascl.net/#1}{\nolinkurl{http://ascl.net/#1}}}
\providecommand{\doarXiv}[1]{\href{https://arxiv.org/abs/#1}{\nolinkurl{https://arxiv.org/abs/#1}}}

\bibitem[{{Agol} {et~al.}(2021){Agol}, {Dorn}, {Grimm}, {Turbet}, {Ducrot},
  {Delrez}, {Gillon}, {Demory}, {Burdanov}, {Barkaoui}, {Benkhaldoun},
  {Bolmont}, {Burgasser}, {Carey}, {de Wit}, {Fabrycky}, {Foreman-Mackey},
  {Haldemann}, {Hernandez}, {Ingalls}, {Jehin}, {Langford}, {Leconte},
  {Lederer}, {Luger}, {Malhotra}, {Meadows}, {Morris}, {Pozuelos}, {Queloz},
  {Raymond}, {Selsis}, {Sestovic}, {Triaud}, \& {Van Grootel}}]{agol2021}
{Agol}, E., {Dorn}, C., {Grimm}, S.~L., {et~al.} 2021, \psj, 2, 1,
  \dodoi{10.3847/PSJ/abd022}

\bibitem[{{Barnes} {et~al.}(2009{\natexlab{a}}){Barnes}, {Jackson},
  {Greenberg}, \& {Raymond}}]{barnes2009b}
{Barnes}, R., {Jackson}, B., {Greenberg}, R., \& {Raymond}, S.~N.
  2009{\natexlab{a}}, \apjl, 700, L30, \dodoi{10.1088/0004-637X/700/1/L30}

\bibitem[{{Barnes} {et~al.}(2009{\natexlab{b}}){Barnes}, {Jackson}, {Raymond},
  {West}, \& {Greenberg}}]{barnes2009a}
{Barnes}, R., {Jackson}, B., {Raymond}, S.~N., {West}, A.~A., \& {Greenberg},
  R. 2009{\natexlab{b}}, \apj, 695, 1006, \dodoi{10.1088/0004-637X/695/2/1006}

\bibitem[{{Bashi} {et~al.}(2020){Bashi}, {Zucker}, {Adibekyan}, {Santos},
  {Tal-Or}, {Trifonov}, \& {Mazeh}}]{bashi2020}
{Bashi}, D., {Zucker}, S., {Adibekyan}, V., {et~al.} 2020, \aap, 643, A106,
  \dodoi{10.1051/0004-6361/202038881}

\bibitem[{{Batygin} \& {Laughlin}(2015)}]{batygin2015b}
{Batygin}, K., \& {Laughlin}, G. 2015, Proceedings of the National Academy of
  Science, 112, 4214, \dodoi{10.1073/pnas.1423252112}

\bibitem[{{Becker} \& {Adams}(2017)}]{becker2017a}
{Becker}, J.~C., \& {Adams}, F.~C. 2017, \mnras, 468, 549,
  \dodoi{10.1093/mnras/stx461}

\bibitem[{{Bolmont} {et~al.}(2016){Bolmont}, {Libert}, {Leconte}, \&
  {Selsis}}]{bolmont2016a}
{Bolmont}, E., {Libert}, A.-S., {Leconte}, J., \& {Selsis}, F. 2016, \aap, 591,
  A106, \dodoi{10.1051/0004-6361/201628073}

\bibitem[{{Bolmont} {et~al.}(2014){Bolmont}, {Raymond}, {von Paris}, {Selsis},
  {Hersant}, {Quintana}, \& {Barclay}}]{bolmont2014}
{Bolmont}, E., {Raymond}, S.~N., {von Paris}, P., {et~al.} 2014, \apj, 793, 3,
  \dodoi{10.1088/0004-637X/793/1/3}

\bibitem[{{Bonfils} {et~al.}(2013){Bonfils}, {Delfosse}, {Udry}, {Forveille},
  {Mayor}, {Perrier}, {Bouchy}, {Gillon}, {Lovis}, {Pepe}, {Queloz}, {Santos},
  {S{\'e}gransan}, \& {Bertaux}}]{bonfils2013a}
{Bonfils}, X., {Delfosse}, X., {Udry}, S., {et~al.} 2013, \aap, 549, A109,
  \dodoi{10.1051/0004-6361/201014704}

\bibitem[{{Bottke} {et~al.}(2005){Bottke}, {Durda}, {Nesvorn{\'y}}, {Jedicke},
  {Morbidelli}, {Vokrouhlick{\'y}}, \& {Levison}}]{bottke2005b}
{Bottke}, W.~F., {Durda}, D.~D., {Nesvorn{\'y}}, D., {et~al.} 2005, \icarus,
  175, 111, \dodoi{10.1016/j.icarus.2004.10.026}

\bibitem[{{Bowler} {et~al.}(2020){Bowler}, {Blunt}, \& {Nielsen}}]{bowler2020a}
{Bowler}, B.~P., {Blunt}, S.~C., \& {Nielsen}, E.~L. 2020, \aj, 159, 63,
  \dodoi{10.3847/1538-3881/ab5b11}

\bibitem[{{Bromley} \& {Kenyon}(2017)}]{bromley2017}
{Bromley}, B.~C., \& {Kenyon}, S.~J. 2017, \aj, 153, 216,
  \dodoi{10.3847/1538-3881/aa6aaa}

\bibitem[{{Bryson} {et~al.}(2021){Bryson}, {Kunimoto}, {Kopparapu}, {Coughlin},
  {Borucki}, {Koch}, {Aguirre}, {Allen}, {Barentsen}, {Batalha}, {Berger},
  {Boss}, {Buchhave}, {Burke}, {Caldwell}, {Campbell}, {Catanzarite},
  {Chandrasekaran}, {Chaplin}, {Christiansen}, {Christensen-Dalsgaard},
  {Ciardi}, {Clarke}, {Cochran}, {Dotson}, {Doyle}, {Duarte}, {Dunham},
  {Dupree}, {Endl}, {Fanson}, {Ford}, {Fujieh}, {Gautier}, {Geary},
  {Gilliland}, {Girouard}, {Gould}, {Haas}, {Henze}, {Holman}, {Howard},
  {Howell}, {Huber}, {Hunter}, {Jenkins}, {Kjeldsen}, {Kolodziejczak},
  {Larson}, {Latham}, {Li}, {Mathur}, {Meibom}, {Middour}, {Morris}, {Morton},
  {Mullally}, {Mullally}, {Pletcher}, {Prsa}, {Quinn}, {Quintana}, {Ragozzine},
  {Ramirez}, {Sanderfer}, {Sasselov}, {Seader}, {Shabram}, {Shporer}, {Smith},
  {Steffen}, {Still}, {Torres}, {Troeltzsch}, {Twicken}, {Uddin}, {Van Cleve},
  {Voss}, {Weiss}, {Welsh}, {Wohler}, \& {Zamudio}}]{bryson2021}
{Bryson}, S., {Kunimoto}, M., {Kopparapu}, R.~K., {et~al.} 2021, \aj, 161, 36,
  \dodoi{10.3847/1538-3881/abc418}

\bibitem[{{Carrera} {et~al.}(2019){Carrera}, {Raymond}, \&
  {Davies}}]{carrera2019b}
{Carrera}, D., {Raymond}, S.~N., \& {Davies}, M.~B. 2019, \aap, 629, L7,
  \dodoi{10.1051/0004-6361/201935744}

\bibitem[{{Chambers}(1999)}]{chambers1999}
{Chambers}, J.~E. 1999, \mnras, 304, 793,
  \dodoi{10.1046/j.1365-8711.1999.02379.x}

\bibitem[{{Chambers} {et~al.}(1996){Chambers}, {Wetherill}, \&
  {Boss}}]{chambers1996}
{Chambers}, J.~E., {Wetherill}, G.~W., \& {Boss}, A.~P. 1996, \icarus, 119,
  261, \dodoi{10.1006/icar.1996.0019}

\bibitem[{{Clement} {et~al.}(2021){Clement}, {Kaib}, {Raymond}, \&
  {Chambers}}]{clement2021f}
{Clement}, M.~S., {Kaib}, N.~A., {Raymond}, S.~N., \& {Chambers}, J.~E. 2021,
  \icarus, 367, 114585, \dodoi{10.1016/j.icarus.2021.114585}

\bibitem[{{Clement} {et~al.}(2018){Clement}, {Kaib}, {Raymond}, \&
  {Walsh}}]{clement2018}
{Clement}, M.~S., {Kaib}, N.~A., {Raymond}, S.~N., \& {Walsh}, K.~J. 2018,
  \icarus, 311, 340, \dodoi{10.1016/j.icarus.2018.04.008}

\bibitem[{{Clement} {et~al.}(2020){Clement}, {Morbidelli}, {Raymond}, \&
  {Kaib}}]{clement2020a}
{Clement}, M.~S., {Morbidelli}, A., {Raymond}, S.~N., \& {Kaib}, N.~A. 2020,
  \mnras, 492, L56, \dodoi{10.1093/mnrasl/slz184}

\bibitem[{{Clement} {et~al.}(2022){Clement}, {Quintana}, \&
  {Quarles}}]{clement2022a}
{Clement}, M.~S., {Quintana}, E.~V., \& {Quarles}, B.~L. 2022, \apj, 928, 91,
  \dodoi{10.3847/1538-4357/ac549e}

\bibitem[{{Clement} {et~al.}(2019){Clement}, {Raymond}, \&
  {Kaib}}]{clement2019a}
{Clement}, M.~S., {Raymond}, S.~N., \& {Kaib}, N.~A. 2019, \aj, 157, 38,
  \dodoi{10.3847/1538-3881/aaf21e}

\bibitem[{{Correia} \& {Laskar}(2004)}]{correia2004}
{Correia}, A. C.~M., \& {Laskar}, J. 2004, \nat, 429, 848,
  \dodoi{10.1038/nature02609}

\bibitem[{{Correia} {et~al.}(2011){Correia}, {Laskar}, {Farago}, \&
  {Bou{\'e}}}]{correia2011}
{Correia}, A. C.~M., {Laskar}, J., {Farago}, F., \& {Bou{\'e}}, G. 2011,
  Celestial Mechanics and Dynamical Astronomy, 111, 105,
  \dodoi{10.1007/s10569-011-9368-9}

\bibitem[{{Deienno} {et~al.}(2016){Deienno}, {Gomes}, {Walsh}, {Morbidelli}, \&
  {Nesvorn{\'y}}}]{deienno2016a}
{Deienno}, R., {Gomes}, R.~S., {Walsh}, K.~J., {Morbidelli}, A., \&
  {Nesvorn{\'y}}, D. 2016, \icarus, 272, 114,
  \dodoi{10.1016/j.icarus.2016.02.043}

\bibitem[{{Deitrick} {et~al.}(2018{\natexlab{a}}){Deitrick}, {Barnes}, {Quinn},
  {Armstrong}, {Charnay}, \& {Wilhelm}}]{deitrick2018a}
{Deitrick}, R., {Barnes}, R., {Quinn}, T.~R., {et~al.} 2018{\natexlab{a}}, \aj,
  155, 60, \dodoi{10.3847/1538-3881/aaa301}

\bibitem[{{Deitrick} {et~al.}(2018{\natexlab{b}}){Deitrick}, {Barnes}, {Bitz},
  {Fleming}, {Charnay}, {Meadows}, {Wilhelm}, {Armstrong}, \&
  {Quinn}}]{deitrick2018b}
{Deitrick}, R., {Barnes}, R., {Bitz}, C., {et~al.} 2018{\natexlab{b}}, \aj,
  155, 266, \dodoi{10.3847/1538-3881/aac214}

\bibitem[{{DeMeo} \& {Carry}(2014)}]{demeo2014b}
{DeMeo}, F.~E., \& {Carry}, B. 2014, \nat, 505, 629,
  \dodoi{10.1038/nature12908}

\bibitem[{{Dermott} \& {Murray}(1981)}]{dermott1981b}
{Dermott}, S.~F., \& {Murray}, C.~D. 1981, \nat, 290, 664,
  \dodoi{10.1038/290664a0}

\bibitem[{{Dressing} \& {Charbonneau}(2013)}]{dressing2013}
{Dressing}, C.~D., \& {Charbonneau}, D. 2013, \apj, 767, 95,
  \dodoi{10.1088/0004-637X/767/1/95}

\bibitem[{{Dressing} {et~al.}(2010){Dressing}, {Spiegel}, {Scharf}, {Menou}, \&
  {Raymond}}]{dressing2010}
{Dressing}, C.~D., {Spiegel}, D.~S., {Scharf}, C.~A., {Menou}, K., \&
  {Raymond}, S.~N. 2010, \apj, 721, 1295, \dodoi{10.1088/0004-637X/721/2/1295}

\bibitem[{{Fang} \& {Margot}(2012)}]{fang2012f}
{Fang}, J., \& {Margot}, J.-L. 2012, \apj, 761, 92,
  \dodoi{10.1088/0004-637X/761/2/92}

\bibitem[{{Fischer} {et~al.}(2016){Fischer}, {Anglada-Escude}, {Arriagada},
  {Baluev}, {Bean}, {Bouchy}, {Buchhave}, {Carroll}, {Chakraborty}, {Crepp},
  {Dawson}, {Diddams}, {Dumusque}, {Eastman}, {Endl}, {Figueira}, {Ford},
  {Foreman-Mackey}, {Fournier}, {F{\H{u}}r{\'e}sz}, {Gaudi}, {Gregory},
  {Grundahl}, {Hatzes}, {H{\'e}brard}, {Herrero}, {Hogg}, {Howard}, {Johnson},
  {Jorden}, {Jurgenson}, {Latham}, {Laughlin}, {Loredo}, {Lovis}, {Mahadevan},
  {McCracken}, {Pepe}, {Perez}, {Phillips}, {Plavchan}, {Prato}, {Quirrenbach},
  {Reiners}, {Robertson}, {Santos}, {Sawyer}, {Segransan}, {Sozzetti},
  {Steinmetz}, {Szentgyorgyi}, {Udry}, {Valenti}, {Wang}, {Wittenmyer}, \&
  {Wright}}]{fischer2016}
{Fischer}, D.~A., {Anglada-Escude}, G., {Arriagada}, P., {et~al.} 2016, \pasp,
  128, 066001, \dodoi{10.1088/1538-3873/128/964/066001}

\bibitem[{{Folkner} {et~al.}(2014){Folkner}, {Williams}, {Boggs}, {Park}, \&
  {Kuchynka}}]{folkner2014}
{Folkner}, W.~M., {Williams}, J.~G., {Boggs}, D.~H., {Park}, R.~S., \&
  {Kuchynka}, P. 2014, Interplanetary Network Progress Report, 42-196, 1

\bibitem[{{Ford}(2014)}]{ford2014}
{Ford}, E.~B. 2014, Proceedings of the National Academy of Science, 111, 12616,
  \dodoi{10.1073/pnas.1304219111}

\bibitem[{{Fulton} {et~al.}(2017){Fulton}, {Petigura}, {Howard}, {Isaacson},
  {Marcy}, {Cargile}, {Hebb}, {Weiss}, {Johnson}, {Morton}, {Sinukoff},
  {Crossfield}, \& {Hirsch}}]{fulton2017}
{Fulton}, B.~J., {Petigura}, E.~A., {Howard}, A.~W., {et~al.} 2017, \aj, 154,
  109, \dodoi{10.3847/1538-3881/aa80eb}

\bibitem[{{Fulton} {et~al.}(2021){Fulton}, {Rosenthal}, {Hirsch}, {Isaacson},
  {Howard}, {Dedrick}, {Sherstyuk}, {Blunt}, {Petigura}, {Knutson}, {Behmard},
  {Chontos}, {Crepp}, {Crossfield}, {Dalba}, {Fischer}, {Henry}, {Kane},
  {Kosiarek}, {Marcy}, {Rubenzahl}, {Weiss}, \& {Wright}}]{fulton2021}
{Fulton}, B.~J., {Rosenthal}, L.~J., {Hirsch}, L.~A., {et~al.} 2021, \apjs,
  255, 14, \dodoi{10.3847/1538-4365/abfcc1}

\bibitem[{{Gaia Collaboration} {et~al.}(2021){Gaia Collaboration}, {Brown},
  {Vallenari}, {Prusti}, {de Bruijne}, {Babusiaux}, {Biermann}, {Creevey},
  {Evans}, {Eyer}, {Hutton}, {Jansen}, {Jordi}, {Klioner}, {Lammers},
  {Lindegren}, {Luri}, {Mignard}, {Panem}, {Pourbaix}, {Randich}, {Sartoretti},
  {Soubiran}, {Walton}, {Arenou}, {Bailer-Jones}, {Bastian}, {Cropper},
  {Drimmel}, {Katz}, {Lattanzi}, {van Leeuwen}, {Bakker}, {Cacciari},
  {Casta{\~n}eda}, {De Angeli}, {Ducourant}, {Fabricius}, {Fouesneau},
  {Fr{\'e}mat}, {Guerra}, {Guerrier}, {Guiraud}, {Jean-Antoine Piccolo},
  {Masana}, {Messineo}, {Mowlavi}, {Nicolas}, {Nienartowicz}, {Pailler},
  {Panuzzo}, {Riclet}, {Roux}, {Seabroke}, {Sordo}, {Tanga}, {Th{\'e}venin},
  {Gracia-Abril}, {Portell}, {Teyssier}, {Altmann}, {Andrae}, {Bellas-Velidis},
  {Benson}, {Berthier}, {Blomme}, {Brugaletta}, {Burgess}, {Busso}, {Carry},
  {Cellino}, {Cheek}, {Clementini}, {Damerdji}, {Davidson}, {Delchambre},
  {Dell'Oro}, {Fern{\'a}ndez-Hern{\'a}ndez}, {Galluccio}, {Garc{\'\i}a-Lario},
  {Garcia-Reinaldos}, {Gonz{\'a}lez-N{\'u}{\~n}ez}, {Gosset}, {Haigron},
  {Halbwachs}, {Hambly}, {Harrison}, {Hatzidimitriou}, {Heiter},
  {Hern{\'a}ndez}, {Hestroffer}, {Hodgkin}, {Holl}, {Jan{\ss}en}, {Jevardat de
  Fombelle}, {Jordan}, {Krone-Martins}, {Lanzafame}, {L{\"o}ffler}, {Lorca},
  {Manteiga}, {Marchal}, {Marrese}, {Moitinho}, {Mora}, {Muinonen}, {Osborne},
  {Pancino}, {Pauwels}, {Petit}, {Recio-Blanco}, {Richards}, {Riello},
  {Rimoldini}, {Robin}, {Roegiers}, {Rybizki}, {Sarro}, {Siopis}, {Smith},
  {Sozzetti}, {Ulla}, {Utrilla}, {van Leeuwen}, {van Reeven}, {Abbas}, {Abreu
  Aramburu}, {Accart}, {Aerts}, {Aguado}, {Ajaj}, {Altavilla}, {{\'A}lvarez},
  {{\'A}lvarez Cid-Fuentes}, {Alves}, {Anderson}, {Anglada Varela}, {Antoja},
  {Audard}, {Baines}, {Baker}, {Balaguer-N{\'u}{\~n}ez}, {Balbinot}, {Balog},
  {Barache}, {Barbato}, {Barros}, {Barstow}, {Bartolom{\'e}}, {Bassilana},
  {Bauchet}, {Baudesson-Stella}, {Becciani}, {Bellazzini}, {Bernet}, {Bertone},
  {Bianchi}, {Blanco-Cuaresma}, {Boch}, {Bombrun}, {Bossini}, {Bouquillon},
  {Bragaglia}, {Bramante}, {Breedt}, {Bressan}, {Brouillet}, {Bucciarelli},
  {Burlacu}, {Busonero}, {Butkevich}, {Buzzi}, {Caffau}, {Cancelliere},
  {C{\'a}novas}, {Cantat-Gaudin}, {Carballo}, {Carlucci}, {Carnerero},
  {Carrasco}, {Casamiquela}, {Castellani}, {Castro-Ginard}, {Castro Sampol},
  {Chaoul}, {Charlot}, {Chemin}, {Chiavassa}, {Cioni}, {Comoretto}, {Cooper},
  {Cornez}, {Cowell}, {Crifo}, {Crosta}, {Crowley}, {Dafonte}, {Dapergolas},
  {David}, {David}, {de Laverny}, {De Luise}, {De March}, {De Ridder}, {de
  Souza}, {de Teodoro}, {de Torres}, {del Peloso}, {del Pozo}, {Delbo},
  {Delgado}, {Delgado}, {Delisle}, {Di Matteo}, {Diakite}, {Diener},
  {Distefano}, {Dolding}, {Eappachen}, {Edvardsson}, {Enke}, {Esquej}, {Fabre},
  {Fabrizio}, {Faigler}, {Fedorets}, {Fernique}, {Fienga}, {Figueras},
  {Fouron}, {Fragkoudi}, {Fraile}, {Franke}, {Gai}, {Garabato},
  {Garcia-Gutierrez}, {Garc{\'\i}a-Torres}, {Garofalo}, {Gavras}, {Gerlach},
  {Geyer}, {Giacobbe}, {Gilmore}, {Girona}, {Giuffrida}, {Gomel}, {Gomez},
  {Gonzalez-Santamaria}, {Gonz{\'a}lez-Vidal}, {Granvik},
  {Guti{\'e}rrez-S{\'a}nchez}, {Guy}, {Hauser}, {Haywood}, {Helmi}, {Hidalgo},
  {Hilger}, {H{\l}adczuk}, {Hobbs}, {Holland}, {Huckle}, {Jasniewicz},
  {Jonker}, {Juaristi Campillo}, {Julbe}, {Karbevska}, {Kervella}, {Khanna},
  {Kochoska}, {Kontizas}, {Kordopatis}, {Korn}, {Kostrzewa-Rutkowska},
  {Kruszy{\'n}ska}, {Lambert}, {Lanza}, {Lasne}, {Le Campion}, {Le Fustec},
  {Lebreton}, {Lebzelter}, {Leccia}, {Leclerc}, {Lecoeur-Taibi}, {Liao},
  {Licata}, {Lindstr{\o}m}, {Lister}, {Livanou}, {Lobel}, {Madrero Pardo},
  {Managau}, {Mann}, {Marchant}, {Marconi}, {Marcos Santos}, {Marinoni},
  {Marocco}, {Marshall}, {Martin Polo}, {Mart{\'\i}n-Fleitas}, {Masip},
  {Massari}, {Mastrobuono-Battisti}, {Mazeh}, {McMillan}, {Messina},
  {Michalik}, {Millar}, {Mints}, {Molina}, {Molinaro}, {Moln{\'a}r},
  {Montegriffo}, {Mor}, {Morbidelli}, {Morel}, {Morris}, {Mulone}, {Munoz},
  {Muraveva}, {Murphy}, {Musella}, {Noval}, {Ord{\'e}novic}, {Orr{\`u}},
  {Osinde}, {Pagani}, {Pagano}, {Palaversa}, {Palicio}, {Panahi}, {Pawlak},
  {Pe{\~n}alosa Esteller}, {Penttil{\"a}}, {Piersimoni}, {Pineau}, {Plachy},
  {Plum}, {Poggio}, {Poretti}, {Poujoulet}, {Pr{\v{s}}a}, {Pulone}, {Racero},
  {Ragaini}, {Rainer}, {Raiteri}, {Rambaux}, {Ramos}, {Ramos-Lerate}, {Re
  Fiorentin}, {Regibo}, {Reyl{\'e}}, {Ripepi}, {Riva}, {Rixon}, {Robichon},
  {Robin}, {Roelens}, {Rohrbasser}, {Romero-G{\'o}mez}, {Rowell}, {Royer},
  {Rybicki}, {Sadowski}, {Sagrist{\`a} Sell{\'e}s}, {Sahlmann}, {Salgado},
  {Salguero}, {Samaras}, {Sanchez Gimenez}, {Sanna}, {Santove{\~n}a},
  {Sarasso}, {Schultheis}, {Sciacca}, {Segol}, {Segovia}, {S{\'e}gransan},
  {Semeux}, {Shahaf}, {Siddiqui}, {Siebert}, {Siltala}, {Slezak}, {Smart},
  {Solano}, {Solitro}, {Souami}, {Souchay}, {Spagna}, {Spoto}, {Steele},
  {Steidelm{\"u}ller}, {Stephenson}, {S{\"u}veges}, {Szabados}, {Szegedi-Elek},
  {Taris}, {Tauran}, {Taylor}, {Teixeira}, {Thuillot}, {Tonello}, {Torra},
  {Torra}, {Turon}, {Unger}, {Vaillant}, {van Dillen}, {Vanel}, {Vecchiato},
  {Viala}, {Vicente}, {Voutsinas}, {Weiler}, {Wevers}, {Wyrzykowski}, {Yoldas},
  {Yvard}, {Zhao}, {Zorec}, {Zucker}, {Zurbach}, \& {Zwitter}}]{brown2021}
{Gaia Collaboration}, {Brown}, A.~G.~A., {Vallenari}, A., {et~al.} 2021, \aap,
  649, A1, \dodoi{10.1051/0004-6361/202039657}

\bibitem[{{Gelino} \& {Kane}(2014)}]{gelino2014}
{Gelino}, D.~M., \& {Kane}, S.~R. 2014, \apj, 787, 105,
  \dodoi{10.1088/0004-637X/787/2/105}

\bibitem[{{Gillon} {et~al.}(2016){Gillon}, {Jehin}, {Lederer}, {Delrez}, {de
  Wit}, {Burdanov}, {Van Grootel}, {Burgasser}, {Triaud}, {Opitom}, {Demory},
  {Sahu}, {Bardalez Gagliuffi}, {Magain}, \& {Queloz}}]{gillon2016}
{Gillon}, M., {Jehin}, E., {Lederer}, S.~M., {et~al.} 2016, \nat, 533, 221,
  \dodoi{10.1038/nature17448}

\bibitem[{{Gillon} {et~al.}(2017){Gillon}, {Triaud}, {Demory}, {Jehin}, {Agol},
  {Deck}, {Lederer}, {de Wit}, {Burdanov}, {Ingalls}, {Bolmont}, {Leconte},
  {Raymond}, {Selsis}, {Turbet}, {Barkaoui}, {Burgasser}, {Burleigh}, {Carey},
  {Chaushev}, {Copperwheat}, {Delrez}, {Fernandes}, {Holdsworth}, {Kotze}, {Van
  Grootel}, {Almleaky}, {Benkhaldoun}, {Magain}, \& {Queloz}}]{gillon2017a}
{Gillon}, M., {Triaud}, A.~H.~M.~J., {Demory}, B.-O., {et~al.} 2017, \nat, 542,
  456, \dodoi{10.1038/nature21360}

\bibitem[{{Gladman} {et~al.}(1997){Gladman}, {Migliorini}, {Morbidelli},
  {Zappala}, {Michel}, {Cellino}, {Froeschle}, {Levison}, {Bailey}, \&
  {Duncan}}]{gladman1997b}
{Gladman}, B.~J., {Migliorini}, F., {Morbidelli}, A., {et~al.} 1997, Science,
  277, 197, \dodoi{10.1126/science.277.5323.197}

\bibitem[{{Gomes} {et~al.}(2005){Gomes}, {Levison}, {Tsiganis}, \&
  {Morbidelli}}]{gomes2005b}
{Gomes}, R., {Levison}, H.~F., {Tsiganis}, K., \& {Morbidelli}, A. 2005, \nat,
  435, 466, \dodoi{10.1038/nature03676}

\bibitem[{{Gupta} {et~al.}(2021){Gupta}, {Wright}, {Robertson}, {Halverson},
  {Luhn}, {Roy}, {Mahadevan}, {Ford}, {Bender}, {Blake}, {Hearty}, {Kanodia},
  {Logsdon}, {McElwain}, {Monson}, {Ninan}, {Schwab}, {Stef{\'a}nsson}, \&
  {Terrien}}]{gupta2021}
{Gupta}, A.~F., {Wright}, J.~T., {Robertson}, P., {et~al.} 2021, \aj, 161, 130,
  \dodoi{10.3847/1538-3881/abd79e}

\bibitem[{{Hansen}(2009)}]{hansen2009c}
{Hansen}, B. M.~S. 2009, \apj, 703, 1131, \dodoi{10.1088/0004-637X/703/1/1131}

\bibitem[{{Hansen}(2017)}]{hansen2017b}
---. 2017, \mnras, 467, 1531, \dodoi{10.1093/mnras/stx182}

\bibitem[{{Hatzes}(2016)}]{hatzes2016d}
{Hatzes}, A.~P. 2016, \ssr, 205, 267, \dodoi{10.1007/s11214-016-0246-3}

\bibitem[{{He} {et~al.}(2019){He}, {Ford}, \& {Ragozzine}}]{he2019}
{He}, M.~Y., {Ford}, E.~B., \& {Ragozzine}, D. 2019, \mnras, 490, 4575,
  \dodoi{10.1093/mnras/stz2869}

\bibitem[{{Horner} {et~al.}(2020{\natexlab{a}}){Horner}, {Vervoort}, {Kane},
  {Ceja}, {Waltham}, {Gilmore}, \& {Kirtland Turner}}]{horner2020a}
{Horner}, J., {Vervoort}, P., {Kane}, S.~R., {et~al.} 2020{\natexlab{a}}, \aj,
  159, 10, \dodoi{10.3847/1538-3881/ab5365}

\bibitem[{{Horner} {et~al.}(2020{\natexlab{b}}){Horner}, {Kane}, {Marshall},
  {Dalba}, {Holt}, {Wood}, {Maynard-Casely}, {Wittenmyer}, {Lykawka}, {Hill},
  {Salmeron}, {Bailey}, {L{\"o}hne}, {Agnew}, {Carter}, \&
  {Tylor}}]{horner2020b}
{Horner}, J., {Kane}, S.~R., {Marshall}, J.~P., {et~al.} 2020{\natexlab{b}},
  \pasp, 132, 102001, \dodoi{10.1088/1538-3873/ab8eb9}

\bibitem[{{Howard} {et~al.}(2010){Howard}, {Marcy}, {Johnson}, {Fischer},
  {Wright}, {Isaacson}, {Valenti}, {Anderson}, {Lin}, \& {Ida}}]{howard2010b}
{Howard}, A.~W., {Marcy}, G.~W., {Johnson}, J.~A., {et~al.} 2010, Science, 330,
  653, \dodoi{10.1126/science.1194854}

\bibitem[{{Izidoro} {et~al.}(2015){Izidoro}, {Raymond}, {Morbidelli}, \&
  {Winter}}]{izidoro2015c}
{Izidoro}, A., {Raymond}, S.~N., {Morbidelli}, A., \& {Winter}, O.~C. 2015,
  \mnras, 453, 3619, \dodoi{10.1093/mnras/stv1835}

\bibitem[{{Izidoro} {et~al.}(2016){Izidoro}, {Raymond}, {Pierens},
  {Morbidelli}, {Winter}, \& {Nesvorny`}}]{izidoro2016}
{Izidoro}, A., {Raymond}, S.~N., {Pierens}, A., {et~al.} 2016, \apj, 833, 40,
  \dodoi{10.3847/1538-4357/833/1/40}

\bibitem[{{Juri{\'c}} \& {Tremaine}(2008)}]{juric2008b}
{Juri{\'c}}, M., \& {Tremaine}, S. 2008, \apj, 686, 603, \dodoi{10.1086/590047}

\bibitem[{{Kaib} \& {Chambers}(2016)}]{kaib2016a}
{Kaib}, N.~A., \& {Chambers}, J.~E. 2016, \mnras, 455, 3561,
  \dodoi{10.1093/mnras/stv2554}

\bibitem[{{Kane}(2016)}]{kane2016d}
{Kane}, S.~R. 2016, \apj, 830, 105, \dodoi{10.3847/0004-637X/830/2/105}

\bibitem[{{Kane}(2019)}]{kane2019c}
---. 2019, \aj, 158, 72, \dodoi{10.3847/1538-3881/ab2a09}

\bibitem[{{Kane} {et~al.}(2012){Kane}, {Ciardi}, {Gelino}, \& {von
  Braun}}]{kane2012d}
{Kane}, S.~R., {Ciardi}, D.~R., {Gelino}, D.~M., \& {von Braun}, K. 2012,
  \mnras, 425, 757, \dodoi{10.1111/j.1365-2966.2012.21627.x}

\bibitem[{{Kane} \& {Gelino}(2012)}]{kane2012e}
{Kane}, S.~R., \& {Gelino}, D.~M. 2012, Astrobiology, 12, 940,
  \dodoi{10.1089/ast.2011.0798}

\bibitem[{{Kane} {et~al.}(2021{\natexlab{a}}){Kane}, {Jansen}, {Fauchez},
  {Selsis}, \& {Ceja}}]{kane2021c}
{Kane}, S.~R., {Jansen}, T., {Fauchez}, T., {Selsis}, F., \& {Ceja}, A.~Y.
  2021{\natexlab{a}}, \aj, 161, 53, \dodoi{10.3847/1538-3881/abcfbe}

\bibitem[{{Kane} {et~al.}(2021{\natexlab{b}}){Kane}, {Li}, {Wolf}, {Ostberg},
  \& {Hill}}]{kane2021a}
{Kane}, S.~R., {Li}, Z., {Wolf}, E.~T., {Ostberg}, C., \& {Hill}, M.~L.
  2021{\natexlab{b}}, \aj, 161, 31, \dodoi{10.3847/1538-3881/abcbfd}

\bibitem[{{Kane} \& {Raymond}(2014)}]{kane2014b}
{Kane}, S.~R., \& {Raymond}, S.~N. 2014, \apj, 784, 104,
  \dodoi{10.1088/0004-637X/784/2/104}

\bibitem[{{Kane} \& {Torres}(2017)}]{kane2017d}
{Kane}, S.~R., \& {Torres}, S.~M. 2017, \aj, 154, 204,
  \dodoi{10.3847/1538-3881/aa8fce}

\bibitem[{{Kane} {et~al.}(2020{\natexlab{a}}){Kane}, {Turnbull}, {Fulton},
  {Rosenthal}, {Howard}, {Isaacson}, {Marcy}, \& {Weiss}}]{kane2020b}
{Kane}, S.~R., {Turnbull}, M.~C., {Fulton}, B.~J., {et~al.} 2020{\natexlab{a}},
  \aj, 160, 81, \dodoi{10.3847/1538-3881/ab9ffe}

\bibitem[{{Kane} {et~al.}(2020{\natexlab{b}}){Kane}, {Vervoort}, {Horner}, \&
  {Pozuelos}}]{kane2020e}
{Kane}, S.~R., {Vervoort}, P., {Horner}, J., \& {Pozuelos}, F.~J.
  2020{\natexlab{b}}, \psj, 1, 42, \dodoi{10.3847/PSJ/abae63}

\bibitem[{{Kane} {et~al.}(2016){Kane}, {Wittenmyer}, {Hinkel}, {Roy},
  {Mahadevan}, {Dragomir}, {Matthews}, {Henry}, {Chakraborty}, {Boyajian},
  {Wright}, {Ciardi}, {Fischer}, {Butler}, {Tinney}, {Carter}, {Jones},
  {Bailey}, \& {O'Toole}}]{kane2016b}
{Kane}, S.~R., {Wittenmyer}, R.~A., {Hinkel}, N.~R., {et~al.} 2016, \apj, 821,
  65, \dodoi{10.3847/0004-637X/821/1/65}

\bibitem[{{Kane} {et~al.}(2021{\natexlab{c}}){Kane}, {Arney}, {Byrne}, {Dalba},
  {Desch}, {Horner}, {Izenberg}, {Mandt}, {Meadows}, \& {Quick}}]{kane2021d}
{Kane}, S.~R., {Arney}, G.~N., {Byrne}, P.~K., {et~al.} 2021{\natexlab{c}},
  Journal of Geophysical Research (Planets), 126, e06643,
  \dodoi{10.1002/jgre.v126.2}

\bibitem[{{Kopparapu}(2013)}]{kopparapu2013b}
{Kopparapu}, R.~K. 2013, \apj, 767, L8, \dodoi{10.1088/2041-8205/767/1/L8}

\bibitem[{{Kunimoto} \& {Matthews}(2020)}]{kunimoto2020b}
{Kunimoto}, M., \& {Matthews}, J.~M. 2020, \aj, 159, 248,
  \dodoi{10.3847/1538-3881/ab88b0}

\bibitem[{{Laskar}(1988)}]{laskar1988b}
{Laskar}, J. 1988, \aap, 198, 341

\bibitem[{{Laskar}(1989)}]{laskar1989}
---. 1989, \nat, 338, 237, \dodoi{10.1038/338237a0}

\bibitem[{{Laskar} {et~al.}(2012){Laskar}, {Bou{\'e}}, \&
  {Correia}}]{laskar2012}
{Laskar}, J., {Bou{\'e}}, G., \& {Correia}, A.~C.~M. 2012, \aap, 538, A105,
  \dodoi{10.1051/0004-6361/201116643}

\bibitem[{{Lee}(2019)}]{lee2019}
{Lee}, E.~J. 2019, \apj, 878, 36, \dodoi{10.3847/1538-4357/ab1b40}

\bibitem[{{L{\'e}ger} {et~al.}(2009){L{\'e}ger}, {Rouan}, {Schneider}, {Barge},
  {Fridlund}, {Samuel}, {Ollivier}, {Guenther}, {Deleuil}, {Deeg}, {Auvergne},
  {Alonso}, {Aigrain}, {Alapini}, {Almenara}, {Baglin}, {Barbieri}, {Bruntt},
  {Bord{\'e}}, {Bouchy}, {Cabrera}, {Catala}, {Carone}, {Carpano}, {Csizmadia},
  {Dvorak}, {Erikson}, {Ferraz-Mello}, {Foing}, {Fressin}, {Gand olfi},
  {Gillon}, {Gondoin}, {Grasset}, {Guillot}, {Hatzes}, {H{\'e}brard}, {Jorda},
  {Lammer}, {Llebaria}, {Loeillet}, {Mayor}, {Mazeh}, {Moutou}, {P{\"a}tzold},
  {Pont}, {Queloz}, {Rauer}, {Renner}, {Samadi}, {Shporer}, {Sotin}, {Tingley},
  {Wuchterl}, {Adda}, {Agogu}, {Appourchaux}, {Ballans}, {Baron}, {Beaufort},
  {Bellenger}, {Berlin}, {Bernardi}, {Blouin}, {Baudin}, {Bodin}, {Boisnard},
  {Boit}, {Bonneau}, {Borzeix}, {Briet}, {Buey}, {Butler}, {Cailleau},
  {Cautain}, {Chabaud}, {Chaintreuil}, {Chiavassa}, {Costes}, {Cuna Parrho},
  {de Oliveira Fialho}, {Decaudin}, {Defise}, {Djalal}, {Epstein}, {Exil},
  {Faur{\'e}}, {Fenouillet}, {Gaboriaud}, {Gallic}, {Gamet}, {Gavalda},
  {Grolleau}, {Gruneisen}, {Gueguen}, {Guis}, {Guivarc'h}, {Guterman},
  {Hallouard}, {Hasiba}, {Heuripeau}, {Huntzinger}, {Hustaix}, {Imad},
  {Imbert}, {Johlander}, {Jouret}, {Journoud}, {Karioty}, {Kerjean},
  {Lafaille}, {Lafond}, {Lam-Trong}, {Landiech}, {Lapeyrere}, {Larqu{\'e}},
  {Laudet}, {Lautier}, {Lecann}, {Lefevre}, {Leruyet}, {Levacher}, {Magnan},
  {Mazy}, {Mertens}, {Mesnager}, {Meunier}, {Michel}, {Monjoin}, {Naudet},
  {Nguyen-Kim}, {Orcesi}, {Ottacher}, {Perez}, {Peter}, {Plasson}, {Plesseria},
  {Pontet}, {Pradines}, {Quentin}, {Reynaud}, {Rolland }, {Rollenhagen},
  {Romagnan}, {Russ}, {Schmidt}, {Schwartz}, {Sebbag}, {Sedes}, {Smit},
  {Steller}, {Sunter}, {Surace}, {Tello}, {Tiph{\`e}ne}, {Toulouse}, {Ulmer},
  {Vandermarcq}, {Vergnault}, {Vuillemin}, \& {Zanatta}}]{leger2009}
{L{\'e}ger}, A., {Rouan}, D., {Schneider}, J., {et~al.} 2009, \aap, 506, 287,
  \dodoi{10.1051/0004-6361/200911933}

\bibitem[{{Limbach} \& {Turner}(2015)}]{limbach2015}
{Limbach}, M.~A., \& {Turner}, E.~L. 2015, Proceedings of the National Academy
  of Science, 112, 20, \dodoi{10.1073/pnas.1406545111}

\bibitem[{{Linsenmeier} {et~al.}(2015){Linsenmeier}, {Pascale}, \&
  {Lucarini}}]{linsenmeier2015}
{Linsenmeier}, M., {Pascale}, S., \& {Lucarini}, V. 2015, \planss, 105, 43,
  \dodoi{10.1016/j.pss.2014.11.003}

\bibitem[{{Lissauer} {et~al.}(2001){Lissauer}, {Quintana}, {Rivera}, \&
  {Duncan}}]{lissauer2001c}
{Lissauer}, J.~J., {Quintana}, E.~V., {Rivera}, E.~J., \& {Duncan}, M.~J. 2001,
  \icarus, 154, 449, \dodoi{10.1006/icar.2001.6692}

\bibitem[{{Lissauer} {et~al.}(2011){Lissauer}, {Fabrycky}, {Ford}, {Borucki},
  {Fressin}, {Marcy}, {Orosz}, {Rowe}, {Torres}, {Welsh}, {Batalha}, {Bryson},
  {Buchhave}, {Caldwell}, {Carter}, {Charbonneau}, {Christiansen}, {Cochran},
  {Desert}, {Dunham}, {Fanelli}, {Fortney}, {Gautier}, {Geary}, {Gilliland},
  {Haas}, {Hall}, {Holman}, {Koch}, {Latham}, {Lopez}, {McCauliff}, {Miller},
  {Morehead}, {Quintana}, {Ragozzine}, {Sasselov}, {Short}, \&
  {Steffen}}]{lissauer2011a}
{Lissauer}, J.~J., {Fabrycky}, D.~C., {Ford}, E.~B., {et~al.} 2011, \nat, 470,
  53, \dodoi{10.1038/nature09760}

\bibitem[{{Lithwick} \& {Wu}(2011)}]{lithwick2011b}
{Lithwick}, Y., \& {Wu}, Y. 2011, \apj, 739, 31,
  \dodoi{10.1088/0004-637X/739/1/31}

\bibitem[{{Lopez} \& {Fortney}(2013)}]{lopez2013}
{Lopez}, E.~D., \& {Fortney}, J.~J. 2013, \apj, 776, 2,
  \dodoi{10.1088/0004-637X/776/1/2}

\bibitem[{{Lopez} \& {Fortney}(2014)}]{lopez2014}
---. 2014, \apj, 792, 1, \dodoi{10.1088/0004-637X/792/1/1}

\bibitem[{{Luger} {et~al.}(2017){Luger}, {Sestovic}, {Kruse}, {Grimm},
  {Demory}, {Agol}, {Bolmont}, {Fabrycky}, {Fernandes}, {Van Grootel},
  {Burgasser}, {Gillon}, {Ingalls}, {Jehin}, {Raymond}, {Selsis}, {Triaud},
  {Barclay}, {Barentsen}, {Howell}, {Delrez}, {de Wit}, {Foreman-Mackey},
  {Holdsworth}, {Leconte}, {Lederer}, {Turbet}, {Almleaky}, {Benkhaldoun},
  {Magain}, {Morris}, {Heng}, \& {Queloz}}]{luger2017b}
{Luger}, R., {Sestovic}, M., {Kruse}, E., {et~al.} 2017, Nature Astronomy, 1,
  0129, \dodoi{10.1038/s41550-017-0129}

\bibitem[{{Lykawka} \& {Ito}(2019)}]{lykawka2019}
{Lykawka}, P.~S., \& {Ito}, T. 2019, \apj, 883, 130,
  \dodoi{10.3847/1538-4357/ab3b0a}

\bibitem[{{Martin} \& {Livio}(2015)}]{martin2015b}
{Martin}, R.~G., \& {Livio}, M. 2015, \apj, 810, 105,
  \dodoi{10.1088/0004-637X/810/2/105}

\bibitem[{{Masuda} {et~al.}(2020){Masuda}, {Winn}, \& {Kawahara}}]{masuda2020a}
{Masuda}, K., {Winn}, J.~N., \& {Kawahara}, H. 2020, \aj, 159, 38,
  \dodoi{10.3847/1538-3881/ab5c1d}

\bibitem[{{Minton} \& {Malhotra}(2011)}]{minton2011}
{Minton}, D.~A., \& {Malhotra}, R. 2011, \apj, 732, 53,
  \dodoi{10.1088/0004-637X/732/1/53}

\bibitem[{{Morbidelli} {et~al.}(2010){Morbidelli}, {Brasser}, {Gomes},
  {Levison}, \& {Tsiganis}}]{morbidelli2010a}
{Morbidelli}, A., {Brasser}, R., {Gomes}, R., {Levison}, H.~F., \& {Tsiganis},
  K. 2010, \aj, 140, 1391, \dodoi{10.1088/0004-6256/140/5/1391}

\bibitem[{{Morbidelli} \& {Giorgilli}(1989)}]{morbidelli1989b}
{Morbidelli}, A., \& {Giorgilli}, A. 1989, Celestial Mechanics and Dynamical
  Astronomy, 47, 173, \dodoi{10.1007/BF00051204}

\bibitem[{{Morbidelli} {et~al.}(2005){Morbidelli}, {Levison}, {Tsiganis}, \&
  {Gomes}}]{morbidelli2005}
{Morbidelli}, A., {Levison}, H.~F., {Tsiganis}, K., \& {Gomes}, R. 2005, \nat,
  435, 462, \dodoi{10.1038/nature03540}

\bibitem[{{Morbidelli} {et~al.}(2012){Morbidelli}, {Lunine}, {O'Brien},
  {Raymond}, \& {Walsh}}]{morbidelli2012a}
{Morbidelli}, A., {Lunine}, J.~I., {O'Brien}, D.~P., {Raymond}, S.~N., \&
  {Walsh}, K.~J. 2012, Annual Review of Earth and Planetary Sciences, 40, 251,
  \dodoi{10.1146/annurev-earth-042711-105319}

\bibitem[{{Morbidelli} \& {Raymond}(2016)}]{morbidelli2016b}
{Morbidelli}, A., \& {Raymond}, S.~N. 2016, Journal of Geophysical Research
  (Planets), 121, 1962, \dodoi{10.1002/2016JE005088}

\bibitem[{{Mulders} {et~al.}(2021){Mulders}, {Dr{{a}}{\.z}kowska}, {van der
  Marel}, {Ciesla}, \& {Pascucci}}]{mulders2021b}
{Mulders}, G.~D., {Dr{{a}}{\.z}kowska}, J., {van der Marel}, N., {Ciesla},
  F.~J., \& {Pascucci}, I. 2021, \apjl, 920, L1,
  \dodoi{10.3847/2041-8213/ac2947}

\bibitem[{{Nesvorn{\'y}}(2018)}]{nesvorny2018c}
{Nesvorn{\'y}}, D. 2018, \araa, 56, 137,
  \dodoi{10.1146/annurev-astro-081817-052028}

\bibitem[{{Nesvorn{\'y}} \& {Morbidelli}(2012)}]{nesvorny2012c}
{Nesvorn{\'y}}, D., \& {Morbidelli}, A. 2012, \aj, 144, 117,
  \dodoi{10.1088/0004-6256/144/4/117}

\bibitem[{{Nesvorn{\'y}} {et~al.}(2021){Nesvorn{\'y}}, {Roig}, \&
  {Deienno}}]{nesvorny2021a}
{Nesvorn{\'y}}, D., {Roig}, F.~V., \& {Deienno}, R. 2021, \aj, 161, 50,
  \dodoi{10.3847/1538-3881/abc8ef}

\bibitem[{{Nielsen} {et~al.}(2020){Nielsen}, {Gandolfi}, {Armstrong},
  {Jenkins}, {Fridlund}, {Santos}, {Dai}, {Adibekyan}, {Luque}, {Steffen},
  {Esposito}, {Meru}, {Sabotta}, {Bolmont}, {Kossakowski}, {Otegi}, {Murgas},
  {Stalport}, {Rodler}, {D{\'\i}az}, {Kurtovic}, {Ricker}, {Vanderspek},
  {Latham}, {Seager}, {Winn}, {Jenkins}, {Allart}, {Almenara}, {Barrado},
  {Barros}, {Bayliss}, {Berdi{\~n}as}, {Boisse}, {Bouchy}, {Boyd}, {Brown},
  {Bryant}, {Burke}, {Cochran}, {Cooke}, {Demangeon}, {D{\'\i}az}, {Dittman},
  {Dorn}, {Dumusque}, {Garc{\'\i}a}, {Gonz{\'a}lez-Cuesta}, {Grziwa},
  {Georgieva}, {Guerrero}, {Hatzes}, {Helled}, {Henze}, {Hojjatpanah}, {Korth},
  {Lam}, {Lillo-Box}, {Lopez}, {Livingston}, {Mathur}, {Mousis}, {Narita},
  {Osborn}, {Palle}, {Rojas}, {Persson}, {Quinn}, {Rauer}, {Redfield},
  {Santerne}, {dos Santos}, {Seidel}, {Sousa}, {Ting}, {Turbet}, {Udry}, {Vand
  erburg}, {Van Eylen}, {Vines}, {Wheatley}, \& {Wilson}}]{nielsen2020b}
{Nielsen}, L.~D., {Gandolfi}, D., {Armstrong}, D.~J., {et~al.} 2020, \mnras,
  492, 5399, \dodoi{10.1093/mnras/staa197}

\bibitem[{{Noyelles} {et~al.}(2014){Noyelles}, {Frouard}, {Makarov}, \&
  {Efroimsky}}]{noyelles2014}
{Noyelles}, B., {Frouard}, J., {Makarov}, V.~V., \& {Efroimsky}, M. 2014,
  \icarus, 241, 26, \dodoi{10.1016/j.icarus.2014.05.045}

\bibitem[{{Owen} \& {Wu}(2013)}]{owen2013a}
{Owen}, J.~E., \& {Wu}, Y. 2013, \apj, 775, 105,
  \dodoi{10.1088/0004-637X/775/2/105}

\bibitem[{{Owen} \& {Wu}(2017)}]{owen2017c}
---. 2017, \apj, 847, 29, \dodoi{10.3847/1538-4357/aa890a}

\bibitem[{{Pepe} {et~al.}(2014){Pepe}, {Molaro}, {Cristiani}, {Rebolo},
  {Santos}, {Dekker}, {M{\'e}gevand}, {Zerbi}, {Cabral}, {Di Marcantonio},
  {Abreu}, {Affolter}, {Aliverti}, {Allende Prieto}, {Amate}, {Avila},
  {Baldini}, {Bristow}, {Broeg}, {Cirami}, {Coelho}, {Conconi}, {Coretti},
  {Cupani}, {D'Odorico}, {De Caprio}, {Delabre}, {Dorn}, {Figueira}, {Fragoso},
  {Galeotta}, {Genolet}, {Gomes}, {Gonz{\'a}lez Hern{\'a}ndez}, {Hughes},
  {Iwert}, {Kerber}, {Landoni}, {Lizon}, {Lovis}, {Maire}, {Mannetta},
  {Martins}, {Monteiro}, {Oliveira}, {Poretti}, {Rasilla}, {Riva}, {Santana
  Tschudi}, {Santos}, {Sosnowska}, {Sousa}, {Span{\'o}}, {Tenegi}, {Toso},
  {Vanzella}, {Viel}, \& {Zapatero Osorio}}]{pepe2014a}
{Pepe}, F., {Molaro}, P., {Cristiani}, S., {et~al.} 2014, Astronomische
  Nachrichten, 335, 8, \dodoi{12.1002/asna.201312004}

\bibitem[{{Pirani} {et~al.}(2019){Pirani}, {Johansen}, {Bitsch}, {Mustill}, \&
  {Turrini}}]{pirani2019a}
{Pirani}, S., {Johansen}, A., {Bitsch}, B., {Mustill}, A.~J., \& {Turrini}, D.
  2019, \aap, 623, A169, \dodoi{10.1051/0004-6361/201833713}

\bibitem[{{Raymond} \& {Izidoro}(2017)}]{raymond2017a}
{Raymond}, S.~N., \& {Izidoro}, A. 2017, Science Advances, 3, e1701138,
  \dodoi{10.1126/sciadv.1701138}

\bibitem[{{Raymond} {et~al.}(2014){Raymond}, {Kokubo}, {Morbidelli},
  {Morishima}, \& {Walsh}}]{raymond2014a}
{Raymond}, S.~N., {Kokubo}, E., {Morbidelli}, A., {Morishima}, R., \& {Walsh},
  K.~J. 2014, in Protostars and Planets VI, ed. H.~{Beuther}, R.~S. {Klessen},
  C.~P. {Dullemond}, \& T.~{Henning}, 595,
  \dodoi{10.2458/azu_uapress_9780816531240-ch026}

\bibitem[{{Raymond} {et~al.}(2009){Raymond}, {O'Brien}, {Morbidelli}, \&
  {Kaib}}]{raymond2009c}
{Raymond}, S.~N., {O'Brien}, D.~P., {Morbidelli}, A., \& {Kaib}, N.~A. 2009,
  \icarus, 203, 644, \dodoi{10.1016/j.icarus.2009.05.016}

\bibitem[{{Rosenthal} {et~al.}(2021){Rosenthal}, {Fulton}, {Hirsch},
  {Isaacson}, {Howard}, {Dedrick}, {Sherstyuk}, {Blunt}, {Petigura}, {Knutson},
  {Behmard}, {Chontos}, {Crepp}, {Crossfield}, {Dalba}, {Fischer}, {Henry},
  {Kane}, {Kosiarek}, {Marcy}, {Rubenzahl}, {Weiss}, \&
  {Wright}}]{rosenthal2021}
{Rosenthal}, L.~J., {Fulton}, B.~J., {Hirsch}, L.~A., {et~al.} 2021, \apjs,
  255, 8, \dodoi{10.3847/1538-4365/abe23c}

\bibitem[{{Sabotta} {et~al.}(2021){Sabotta}, {Schlecker}, {Chaturvedi},
  {Guenther}, {Mu{\~n}oz Rodr{\'\i}guez}, {Mu{\~n}oz S{\'a}nchez}, {Caballero},
  {Shan}, {Reffert}, {Ribas}, {Reiners}, {Hatzes}, {Amado}, {Klahr}, {Morales},
  {Quirrenbach}, {Henning}, {Dreizler}, {Pall{\'e}}, {Perger}, {Azzaro},
  {Jeffers}, {Kaminski}, {K{\"u}rster}, {Lafarga}, {Montes}, {Passegger}, \&
  {Zechmeister}}]{sabotta2021}
{Sabotta}, S., {Schlecker}, M., {Chaturvedi}, P., {et~al.} 2021, \aap, 653,
  A114, \dodoi{10.1051/0004-6361/202140968}

\bibitem[{{Shen} \& {Turner}(2008)}]{shen2008c}
{Shen}, Y., \& {Turner}, E.~L. 2008, \apj, 685, 553, \dodoi{10.1086/590548}

\bibitem[{{Spalding}(2018)}]{spalding2018c}
{Spalding}, C. 2018, \apjl, 869, L17, \dodoi{10.3847/2041-8213/aaf478}

\bibitem[{{Stefansson} {et~al.}(2016){Stefansson}, {Hearty}, {Robertson},
  {Mahadevan}, {Anderson}, {Levi}, {Bender}, {Nelson}, {Monson}, {Blank},
  {Halverson}, {Henderson}, {Ramsey}, {Roy}, {Schwab}, \&
  {Terrien}}]{stefansson2016}
{Stefansson}, G., {Hearty}, F., {Robertson}, P., {et~al.} 2016, \apj, 833, 175,
  \dodoi{10.3847/1538-4357/833/2/175}

\bibitem[{{Tremaine}(2015)}]{tremaine2015}
{Tremaine}, S. 2015, \apj, 807, 157, \dodoi{10.1088/0004-637X/807/2/157}

\bibitem[{{Tremaine} \& {Dong}(2012)}]{tremaine2012}
{Tremaine}, S., \& {Dong}, S. 2012, \aj, 143, 94,
  \dodoi{10.1088/0004-6256/143/4/94}

\bibitem[{{Tsiganis} {et~al.}(2005){Tsiganis}, {Gomes}, {Morbidelli}, \&
  {Levison}}]{tsiganis2005b}
{Tsiganis}, K., {Gomes}, R., {Morbidelli}, A., \& {Levison}, H.~F. 2005, \nat,
  435, 459, \dodoi{10.1038/nature03539}

\bibitem[{{Udry} {et~al.}(2019){Udry}, {Dumusque}, {Lovis}, {S{\'e}gransan},
  {Diaz}, {Benz}, {Bouchy}, {Coffinet}, {Lo Curto}, {Mayor}, {Mordasini},
  {Motalebi}, {Pepe}, {Queloz}, {Santos}, {Wyttenbach}, {Alonso}, {Collier
  Cameron}, {Deleuil}, {Figueira}, {Gillon}, {Moutou}, {Pollacco}, \&
  {Pompei}}]{udry2019}
{Udry}, S., {Dumusque}, X., {Lovis}, C., {et~al.} 2019, \aap, 622, A37,
  \dodoi{10.1051/0004-6361/201731173}

\bibitem[{{Valencia} {et~al.}(2007){Valencia}, {Sasselov}, \&
  {O'Connell}}]{valencia2007b}
{Valencia}, D., {Sasselov}, D.~D., \& {O'Connell}, R.~J. 2007, \apj, 665, 1413,
  \dodoi{10.1086/519554}

\bibitem[{{Van Eylen} \& {Albrecht}(2015)}]{vaneylen2015}
{Van Eylen}, V., \& {Albrecht}, S. 2015, \apj, 808, 126,
  \dodoi{10.1088/0004-637X/808/2/126}

\bibitem[{{Veras} \& {Armitage}(2004)}]{veras2004b}
{Veras}, D., \& {Armitage}, P.~J. 2004, \icarus, 172, 349,
  \dodoi{10.1016/j.icarus.2004.06.012}

\bibitem[{{Volk} \& {Gladman}(2015)}]{volk2015}
{Volk}, K., \& {Gladman}, B. 2015, \apjl, 806, L26,
  \dodoi{10.1088/2041-8205/806/2/L26}

\bibitem[{{Walsh} {et~al.}(2011){Walsh}, {Morbidelli}, {Raymond}, {O'Brien}, \&
  {Mandell}}]{walsh2011c}
{Walsh}, K.~J., {Morbidelli}, A., {Raymond}, S.~N., {O'Brien}, D.~P., \&
  {Mandell}, A.~M. 2011, \nat, 475, 206, \dodoi{10.1038/nature10201}

\bibitem[{{Ward} {et~al.}(1976){Ward}, {Colombo}, \& {Franklin}}]{ward1976}
{Ward}, W.~R., {Colombo}, G., \& {Franklin}, F.~A. 1976, \icarus, 28, 441,
  \dodoi{10.1016/0019-1035(76)90117-2}

\bibitem[{{Way} \& {Georgakarakos}(2017)}]{way2017a}
{Way}, M.~J., \& {Georgakarakos}, N. 2017, \apj, 835, L1,
  \dodoi{10.3847/2041-8213/835/1/L1}

\bibitem[{{Williams} \& {Pollard}(2002)}]{williams2002}
{Williams}, D.~M., \& {Pollard}, D. 2002, International Journal of
  Astrobiology, 1, 61, \dodoi{10.1017/S1473550402001064}

\bibitem[{{Winn} \& {Fabrycky}(2015)}]{winn2015}
{Winn}, J.~N., \& {Fabrycky}, D.~C. 2015, \araa, 53, 409,
  \dodoi{10.1146/annurev-astro-082214-122246}

\bibitem[{{Wisdom}(1983)}]{wisdom1983a}
{Wisdom}, J. 1983, \icarus, 56, 51, \dodoi{10.1016/0019-1035(83)90127-6}

\bibitem[{{Wisdom}(2006)}]{wisdom2006b}
---. 2006, \aj, 131, 2294, \dodoi{10.1086/500829}

\bibitem[{{Wisdom} \& {Holman}(1991)}]{wisdom1991}
{Wisdom}, J., \& {Holman}, M. 1991, \aj, 102, 1528, \dodoi{10.1086/115978}

\bibitem[{{Wittenmyer} {et~al.}(2016){Wittenmyer}, {Butler}, {Tinney},
  {Horner}, {Carter}, {Wright}, {Jones}, {Bailey}, \&
  {O'Toole}}]{wittenmyer2016c}
{Wittenmyer}, R.~A., {Butler}, R.~P., {Tinney}, C.~G., {et~al.} 2016, \apj,
  819, 28, \dodoi{10.3847/0004-637X/819/1/28}

\bibitem[{{Wittenmyer} {et~al.}(2020){Wittenmyer}, {Butler}, {Horner}, {Clark},
  {Tinney}, {Carter}, {Wang}, {Johnson}, \& {Collins}}]{wittenmyer2020a}
{Wittenmyer}, R.~A., {Butler}, R.~P., {Horner}, J., {et~al.} 2020, \mnras, 491,
  5248, \dodoi{10.1093/mnras/stz3378}

\bibitem[{{Zechmeister} {et~al.}(2013){Zechmeister}, {K{\"u}rster}, {Endl}, {Lo
  Curto}, {Hartman}, {Nilsson}, {Henning}, {Hatzes}, \&
  {Cochran}}]{zechmeister2013}
{Zechmeister}, M., {K{\"u}rster}, M., {Endl}, M., {et~al.} 2013, \aap, 552,
  A78, \dodoi{10.1051/0004-6361/201116551}

\end{thebibliography}


\end{document}